\documentclass{book}
\usepackage{roboto}

\usepackage[english]{babel}
\usepackage[letterpaper,top=2cm,bottom=2cm,left=3cm,right=3cm,marginparwidth=1.75cm]{geometry}

\usepackage{pdflscape}
\usepackage{lscape}

\usepackage{tabularx}

\usepackage{rotating}
\usepackage{longtable}
\usepackage{array}
\usepackage{float}
\usepackage{amsmath}
\usepackage{graphicx}
\usepackage{inconsolata}
\usepackage[colorlinks=true, allcolors=blue]{hyperref}
\usepackage{enumitem}
\usepackage{epigraph}  % For block quotes

\usepackage{tikz} % This line includes the tikz package

\usepackage{listings}
\usepackage{xcolor}

\usepackage[T1]{fontenc}
\usepackage{inconsolata}  % Inconsolata font for monospace

\definecolor{bgcolor}{rgb}{0.97,0.97,0.97}
\definecolor{codeblue}{rgb}{0.1,0.1,0.8}
\definecolor{codegreen}{rgb}{0,0.4,0}
\definecolor{codegray}{rgb}{0.4,0.4,0.4}
\definecolor{codepurple}{rgb}{0.5,0,0.5}
\definecolor{codered}{rgb}{0.6,0.2,0.2}
\definecolor{lightgray}{rgb}{0.9,0.9,0.9}
\definecolor{darkgray}{rgb}{0.6,0.6,0.6} % Darker gray for Python frames

\makeatletter
\renewcommand{\paragraph}{%
  \@startsection{paragraph}{4}{\z@}{1ex}{-1em}{\normalfont\normalsize\bfseries\color{gray}}}
\makeatother

\usepackage{enumitem}  % Required for custom list settings
\newenvironment{formatteditem}
{
    \begin{itemize}[leftmargin=1cm, labelsep=0.5cm, itemsep=2pt]  % Set up bullet list formatting
}
{
    \end{itemize}  % End of the bullet list
    \vspace{4pt}  % Add space after the list
}

\lstdefinestyle{python}{
    language=Python,
    basicstyle=\ttfamily\small\color{black}\usefont{T1}{zi4}{m}{n},  % Inconsolata for code
    keywordstyle=\bfseries\color{codeblue},  % Bold keywords
    stringstyle=\color{codegreen},  % Strings in green
    commentstyle=\slshape\color{codegray},  % Comments in gray and slanted (not italics)
    showstringspaces=false,
    numbers=left,
    numberstyle=\tiny\color{codegray},  % Line numbers in tiny gray
    stepnumber=1,
    numbersep=8pt,
    frame=single,
    rulecolor=\color{darkgray},  % Darker frame for Python code
    breaklines=true,
    backgroundcolor=\color{bgcolor},
    tabsize=4,
    captionpos=b,
    morekeywords={self, with, as}, % Add more keywords if needed
}

\lstdefinestyle{text}{
    language=,
    basicstyle=\ttfamily\small\color{black}\usefont{T1}{zi4}{m}{n},  % Inconsolata
    stringstyle=\color{codered},
    commentstyle=\color{codegray},
    showstringspaces=false,
    numbers=none,
    frame=single,
    rulecolor=\color{lightgray},  % Lighter frame for text output
    frameround=tttt,
    breaklines=true,
    backgroundcolor=\color{bgcolor},
    tabsize=4,
    captionpos=b,
}

\lstdefinestyle{cmd}{
    language=bash,
    basicstyle=\ttfamily\small\color{black}\usefont{T1}{zi4}{m}{n},  % Inconsolata
    keywordstyle=\bfseries\color{blue},
    stringstyle=\color{codegreen},
    commentstyle=\itshape\color{gray},
    showstringspaces=false,
    numbers=none,
    frame=single,
    rulecolor=\color{darkgray},  % Darker frame for cmd
    breaklines=true,
    backgroundcolor=\color{bgcolor},
    tabsize=4,
    captionpos=b,
}

\lstdefinestyle{cpp}{
    language=C++,  % Change language to C++
    basicstyle=\ttfamily\small\color{black}\usefont{T1}{zi4}{m}{n},  % Inconsolata for code
    keywordstyle=\bfseries\color{codeblue},  % Bold keywords
    stringstyle=\color{codegreen},  % Strings in green
    commentstyle=\slshape\color{codegray},  % Comments in gray and slanted
    showstringspaces=false,
    numbers=left,
    numberstyle=\tiny\color{codegray},  % Line numbers in tiny gray
    stepnumber=1,
    numbersep=8pt,
    frame=single,
    rulecolor=\color{darkgray},  % Darker frame for C++ code
    breaklines=true,
    backgroundcolor=\color{bgcolor},
    tabsize=4,
    captionpos=b,
    morekeywords={class, public, private, protected, virtual, override},  % Add more C++ keywords if needed
}

\title{Deep Learning and Machine Learning with GPGPU and CUDA: Unlocking the Power of Parallel Computing}
\author{
    Ming Li \textsuperscript{*} \\ 
    \textit{Georgia Institute of Technology} \\
    mli694@gatech.edu
    \and
    Ziqian Bi\textsuperscript{*$\dagger$}\\
    \textit{Indiana University} \\
    bizi@iu.edu
    \and
    Tianyang Wang\textsuperscript{*} \\ 
    \textit{Xi’an Jiaotong-Liverpool University} \\
    Tianyang.Wang21@student.xjtlu.edu.cn
    \and
    Yizhu Wen\textsuperscript{*} \\
    \textit{University of Hawaii} \\
    yizhuw@hawaii.edu
    \and
    Qian Niu \\ 
    \textit{Kyoto University} \\
    niu.qian.f44@kyoto-u.jp
    \and
    Xinyuan Song \\
    \textit{Emory University} \\
    xsong30@emory.edu
    \and
    Zekun Jiang \\
    \textit{Sichuan University} \\
    zekun\_jiang@163.com
    \and    
    Junyu Liu \\ 
    \textit{Kyoto University} \\
    liu.junyu.82w@st.kyoto-u.ac.jp
    \and
    Benji Peng \\ 
    \textit{AppCubic} \\
    benji@appcubic.com
    \and
    Sen Zhang \\ 
    \textit{Rutgers University} \\
    sen.z@rutgers.edu
    \and
    Xuanhe Pan \\ 
    \textit{University of Wisconsin-Madison} \\
    xpan73@wisc.edu
    \and
    Jiawei Xu \\ 
    \textit{Purdue University} \\
    xu1644@purdue.edu
    \and
    Jinlang Wang \\ 
    \textit{University of Wisconsin-Madison} \\
    jinlang.wang@wisc.edu
    \and
    Keyu Chen\\ 
    \textit{Georgia Institute of Technology} \\
    kchen637@gatech.edu
    \and
    Caitlyn Heqi Yin \\
    \textit{University of Wisconsin-Madison} \\
    hyin66@wisc.edu
    \and
    Pohsun Feng \\
    \textit{National Taiwan Normal University} \\
    41075018h@ntnu.edu.tw
    \and
    Ming Liu\textsuperscript{$\dagger$} \\ 
    \textit{Purdue University} \\
    liu3183@purdue.edu
}
\date{}

\begin{document}

\maketitle

\epigraph{"Never stop asking questions and seeking answers. Curiosity fuels progress."}{\textit{Jensen Huang}}

\epigraph{"The most powerful technologies are the ones that empower others."}{\textit{Jensen Huang}}

\epigraph{"The biggest risk is not taking any risk."}{\textit{Lisa Su}}

\epigraph{"Only the Paranoid Survive."}{\textit{Andy Glove}}

\tableofcontents  % Generate a Catalog
\cleardoublepage

\setcounter{part}{4}
\part{Mastering GPGPU with CUDA: Unlocking the Power of Parallel Computing}

\setcounter{chapter}{110}
\chapter{Introduction to CPU and GPU}

\section{Overview of Processing Units}

In modern computing, several types of processing units are used to handle various computational tasks. Two of the most common are the Central Processing Unit (CPU) and the Graphics Processing Unit (GPU) \cite{bernaschi2011combined,dally2021evolution,nikolic2022survey}. Each plays a unique role in how a computer operates and is optimized for specific kinds of woCPU is often described as the "brain" of the computer. It handles general-purpose tasks and is designed to execute complex sequences of instructions efficiently \cite{harun2021review}. This makes it perfect for managing different programs, user interactions, and tasks that require precision and logic. Whether running your operating system, managing a spreadsheet, or browsing the web, the CPU excels in handling a wide variety of operations that demand quick thinking and multitasking \cite{harki2020cpu,henning2000spec}.

The GPU, initially built for rendering images and videos, has expanded its role significantly \cite{mittal2014survey}. Today, it’s not only used for graphics but also for processing tasks that require handling large amounts of data at once \cite{oh2004gpu}. Its ability to perform many calculations simultaneously makes it highly effective for areas like machine learning, scientific research, and cryptocurrency mining, where rapid data processing is key \cite{brodtkorb2013graphics,eklund2013medical,bridges2016understanding}.

In addition to CPUs and GPUs, other types of processing units also play specialized roles. Field Programmable Gate Arrays (FPGAs) \cite{asano2009performance} are flexible and can be customized to perform specific tasks, making them valuable in applications that demand real-time performance, such as telecommunications and AI. Application-Specific Integrated Circuits (ASICs) \cite{zuchowski2002hybrid}, on the other hand, are designed for a single purpose and excel in tasks like cryptocurrency mining, but lack flexibility. Digital Signal Processors (DSPs) \cite{eyre2000evolution} are optimized for real-time data tasks, often found in mobile devices and audio processing.

Though these specialized units like FPGAs, ASICs, and DSPs have their uses, the majority of modern computing revolves around CPUs and GPUs. For the rest of this content, we will focus primarily on exploring the architecture and functionality of CPUs and GPUs, along with their respective strengths and common use cases in everyday computing and advanced computational tasks.

% Add photos to represent FPGA, ASIC, DSPs 

\section{Key Differences Between CPUs and GPUs}

The distinction between CPUs and GPUs primarily lies in their design philosophy and the type of tasks they are optimized to handle \cite{rahmad2011comparison}. Below, we will break down the key differences and why each is suited to its specific role.

\subsection{CPU: General-Purpose Computing}

The CPU is a general-purpose processor that is optimized for single-thread performance and low-latency operations. It is composed of a few powerful cores that are capable of handling a wide variety of tasks, one at a time, in quick succession. This makes the CPU ideal for tasks that require high performance in sequential operations or require the management of multiple tasks, such as:

\begin{formatteditem}
    \item Running operating system processes
    \item Executing complex logic
    \item Managing input/output operations
    \item Performing calculations in everyday applications like word processing or web browsing
\end{formatteditem}

\textbf{Example: Sequential operations in Python}

Consider the following Python code that demonstrates how a CPU would handle a series of sequential operations, such as iterating through a list and performing a calculation on each item. Since CPUs are optimized for single-threaded operations, this is a typical example of the type of task where they excel.

\begin{lstlisting}[style=python]
# Example: Sequential CPU task
numbers = [1, 2, 3, 4, 5]
squared_numbers = []

for number in numbers:
  Forquared_numbers.append(number ** 2)

print(squared_numbers)  # Output: [1, 4, 9, 16, 25]
\end{lstlisting}

In this case, the CPU performs each iteration of the loop one after the other in a linear sequence, quickly handling each task.

\subsection{GPU: Specialized for Parallelism}

In contrast to the CPU, the GPU is designed to excel at handling highly parallel tasks. It is composed of thousands of smaller, simpler cores that can perform the same operation on multiple pieces of data simultaneously. This makes the GPU particularly effective for tasks such as:

\begin{itemize}
    \item Image rendering and graphics processing
    \item Machine learning model training
    \item Large-scale scientific simulations
\end{itemize}

\textbf{Example 1: Parallel operations in Python using a GPU library}

While the following example demonstrates the use of a GPU to perform parallel operations, it requires a library like NumPy or CuPy that can offload tasks to a GPU. Let’s look at a simple example of how we might use CuPy (a GPU-accelerated version of NumPy) \cite{nishino2017cupy} to perform parallel matrix operations.

\begin{lstlisting}[style=python]
import cupy as cp

# Create a large matrix
matrix = cp.random.rand(1000, 1000)

# Perform a matrix multiplication (parallelized on the GPU)
result = cp.dot(matrix, matrix)

print(result)
\end{lstlisting}

In this case, the GPU can perform the matrix multiplication much as a CPU could, as the computation is distributed across thousands of cores.

\textbf{Example 2: Parallel operations in Python using TensorFlow}

In this example, we will demonstrate how to use TensorFlow \cite{developers2022tensorflow} to perform parallel matrix operations on a GPU. TensorFlow automatically detects available GPUs and offloads operations to them.

\begin{lstlisting}[style=python]
import tensorflow as tf

# Create a large matrix
matrix = tf.random.uniform((1000, 1000))

# Perform a matrix multiplication (parallelized on the GPU)
result = tf.matmul(matrix, matrix)

# Start a TensorFlow session to execute the operation
tf.print(result)
\end{lstlisting}

In this case, TensorFlow automatically uses the GPU to accelerate the matrix multiplication.

\textbf{Example 3: Parallel operations in Python using PyTorch and GPU}

This example demonstrates the use of a GPU to perform parallel operations using PyTorch \cite{paszke2019pytorch}, a popular deep-learning framework that provides GPU acceleration. Indeep-learning, we perform matrix operations using PyTorch's CUDA support to leverage the GPU.

\begin{lstlisting}[style=python]
import torch

# Check if GPU is available
device = torch.device('cuda' if torch.cuda.is_available() else 'cpu')

# Create a large matrix on the GPU
matrix = torch.rand(1000, 1000, device=device)

# Perform a matrix multiplication (parallelized on the GPU)
result = torch.matmul(matrix, matrix)

print(result)
\end{lstlisting}

In this case, PyTorch automatically takes advantage of the available GPU resources to perform the matrix multiplication in parallel, offering a significant speedup compared to CPU-based computations.

\section{Applications of CPUs vs. GPUs in Modern Computing}

Both CPUs and GPUs have specific strengths that make them suitable for different types of applications in modern computing. Understanding when to use one over the other is key to optimizing the performance of your programs and systems.

\subsection{CPUs in Modern Computing}

CPUs are best suited for general-purpose tasks that require the coordination of various processes and logic. Some examples of applications that are best handled by a CPU include:

\begin{itemize}
    \item Running system-level software (e.g., operating systems, file systems)
    \item Handling applications that require user interaction, like text editors, web browsers, and office suites
    \item Managing background tasks such as scheduling, system monitoring, and communication between devices
\end{itemize}

CPUs also excel in environments where task switching and multitasking are important. For instance, when running multiple applications on a personal computer, the CPU can quickly switch between tasks and allocate resources accordingly.

\subsection{GPUs in Modern Computing}

GPUs, by contrast, are particularly powerful in applications that involve large-scale parallel data processing. Some of the most common use cases for GPUs today include:

\begin{itemize}
    \item \textbf{Graphics rendering}: GPUs were originally designed to handle the demands of rendering high-quality images in video games and simulations.
    \item \textbf{Machine learning}: The field of deep learning relies heavily on GPUs to train complex neural networks, as they can quickly process the large datasets required for training.
    \item \textbf{Scientific computing}: GPUs are also employed in scientific research to perform simulations and calculations at scale, such as weather prediction, molecular modeling, and high-energy physics.
\end{itemize}

\textbf{Example 1: GPU in Matrix Operations}

For instance, in a machine learning context, GPUs are often used to train models that can recognize images or understand natural language. Below is an example using the PyTorch library to demonstrate how GPUs can accelerate training:

\begin{lstlisting}[style=python]
import torch

# Check if GPU is available
device = torch.device('cuda' if torch.cuda.is_available() else 'cpu')

# Sample tensor
data = torch.randn(1000, 1000).to(device)

# Perform a tensor operation
result = data * data

print(result)
\end{lstlisting}

In this example, if a GPU is available, the tensor operations will be performed on it, speeding up the computation.

\textbf{Example 2: GPU in XGBoost}

For instance, in a machine learning context, GPUs are often used to accelerate model training in libraries like XGBoost \cite{chen2016xgboost}. Below is an example using the XGBoost library to demonstrate how GPUs can accelerate training:

\begin{lstlisting}[style=python]
import xgboost as xgb
from sklearn.datasets import load_breast_cancer
from sklearn.model_selection import train_test_split

# Load dataset
data = load_breast_cancer()
X_train, X_test, y_train, y_test = train_test_split(data.data, data.target, test_size=0.2, random_state=42)

# Convert to DMatrix for XGBoost
dtrain = xgb.DMatrix(X_train, label=y_train)
dtest = xgb.DMatrix(X_test, label=y_test)

# Set parameters for GPU usage
params = {
    'max_depth': 3,
    'eta': 0.1,
    'objective': 'binary:logistic',
    'tree_method': 'gpu_hist'  # Use GPU for training
}

# Train the model
bst = xgb.train(params, dtrain, num_boost_round=10)

# Make predictions
preds = bst.predict(dtest)

print(preds)
\end{lstlisting}

In this example, the 'gpu\_hist' tree method allows XGBoost to use the GPU for training, significantly speeding up the process.

\textbf{Example 3: GPU in TensorFlow - Deep Learning}

In this example, we will use TensorFlow to train a dense neural network on the MNIST dataset \cite{xiao2017fashion}. If a GPU is available, the computations will be executed on it, leading to faster training.

\begin{lstlisting}[style=python]
import tensorflow as tf
from tensorflow.keras import layers, models
from tensorflow.keras.datasets import mnist

# Check if GPU is available
physical_devices = tf.config.list_physical_devices('GPU')
if len(physical_devices) > 0:
    print("GPU is available")
else:
    print("GPU not found, using CPU")

# Load MNIST dataset
(train_images, train_labels), (test_images, test_labels) = mnist.load_data()

# Preprocess data
train_images = train_images.reshape((60000, 28 * 28)).astype('float32') / 255
test_images = test_images.reshape((10000, 28 * 28)).astype('float32') / 255

# Define a simple dense model
model = models.Sequential()
model.add(layers.Dense(512, activation='relu', input_shape=(28 * 28,)))
model.add(layers.Dense(10, activation='softmax'))

# Compile the model
model.compile(optimizer='adam',
              loss='sparse_categorical_crossentropy',
              metrics=['accuracy'])

# Train the model
model.fit(train_images, train_labels, epochs=5, batch_size=128)

# Evaluate the model
test_loss, test_acc = model.evaluate(test_images, test_labels)
print(f"Test accuracy: {test_acc}")
\end{lstlisting}

In this example, TensorFlow automatically utilizes available GPU resources for faster training.

\textbf{Example 4: Large-scale 3D visualization using Mayavi with GPU Acceleration}

In this example, we use the Mayavi \cite{ramachandran2011mayavi} to visualize a large 3D scalar field. Mayavi can utilize GPU acceleration for rendering, especially when dealing with large datasets. The code below demonstrates the generation and visualization of a 3D scalar field, which requires significant computational resources for rendering.

\begin{lstlisting}[style=python]
from mayavi import mlab
import numpy as np

# Generate a large 3D scalar field
x, y, z = np.mgrid[-50:50:100j, -50:50:100j, -50:50:100j]
scalars = np.sin(x*y*z) / (x*y*z)

# Visualize the scalar field using Mayavi
mlab.figure(size=(800, 600), bgcolor=(0, 0, 0))
src = mlab.pipeline.scalar_field(x, y, z, scalars)

# Use GPU accelerated volume rendering
mlab.pipeline.volume(src)

# Display the visualization
mlab.show()
\end{lstlisting}

\begin{figure}[H]
    \centering
    \includegraphics[width=0.6\textwidth]{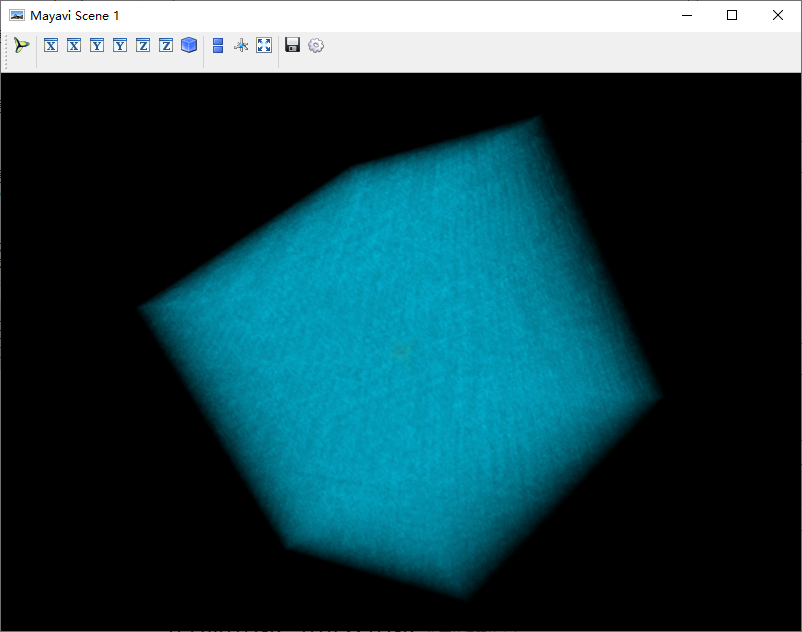}
    \caption{GPUs perform parallel graphics computing}
\end{figure}

In this case, the \texttt{mlab.pipeline.volume} function allows the use of GPU acceleration for rendering the 3D volume, especially when working with larger grids like the one in this example. The larger the grid (100x100x100 here), the more computationally demanding the task becomes, which highlights the benefits of using GPU for rendering.

\subsection{Architecture Comparison}

The CPU and GPU architectures differ fundamentally in their design and purpose. While CPUs have fewer cores, each core is highly sophisticated and capable of handling complex instructions. This makes CPUs ideal for managing general-purpose tasks, with the control unit acting as a ``leader'', coordinating the system. On the other hand, GPUs are equipped with a large number of simple, lightweight cores that excel at parallel processing. The GPU architecture is designed for handling large amounts of simple, repetitive tasks, functioning more like ``workers'' in a large team, efficiently executing multiple tasks simultaneously.

\begin{figure}[H]
    \centering
    \includegraphics[width=0.6\textwidth]{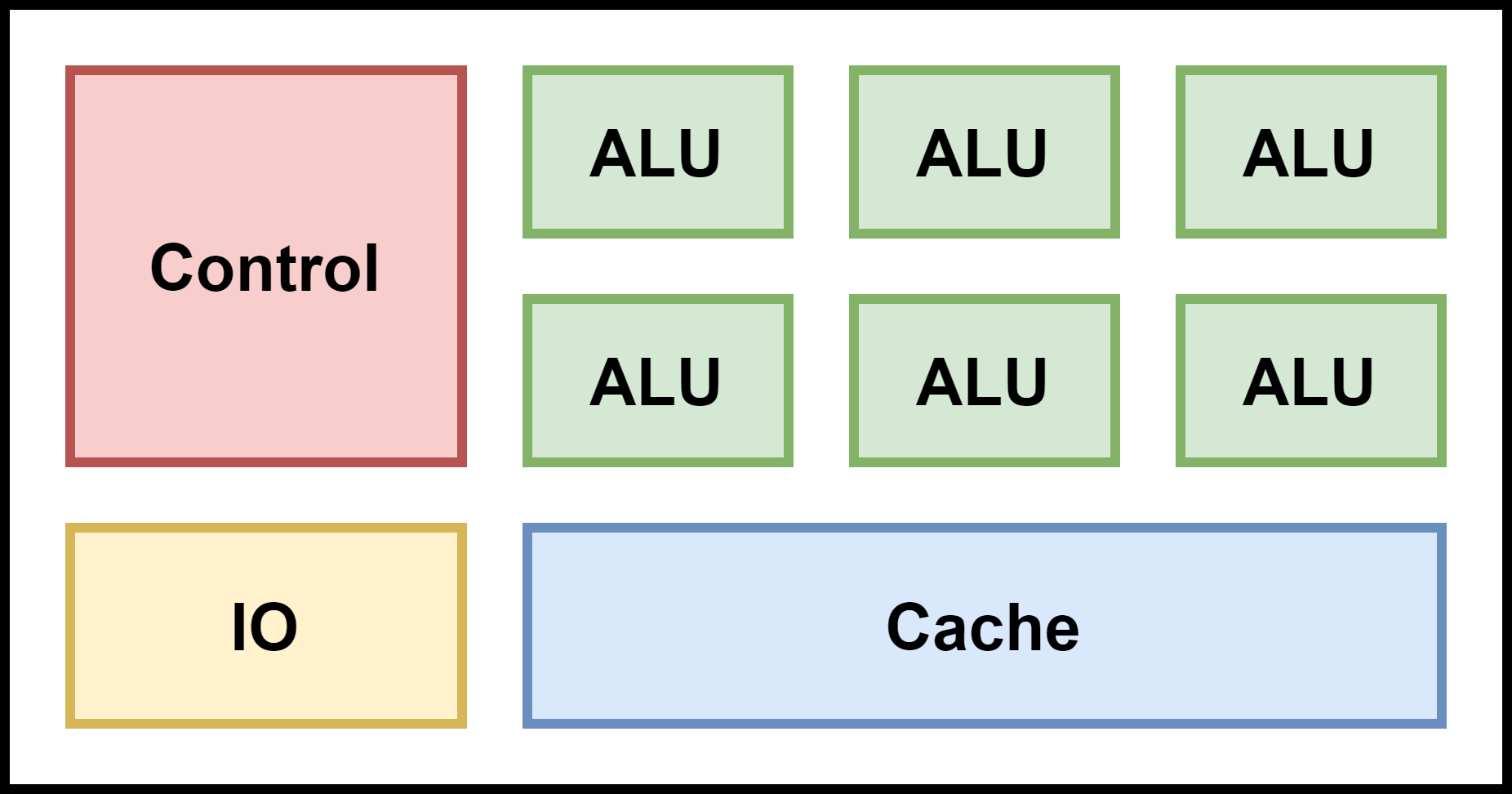}
    \caption{CPU Architecture}
\end{figure}

In the CPU diagram, the control unit coordinates the smaller number of Arithmetic Logic Units (ALUs) to perform general-purpose computation. The IO and cache systems support data transfer and storage, enabling the CPU to handle a wide range of complex tasks.

\begin{figure}[H]
    \centering
    \includegraphics[width=0.6\textwidth]{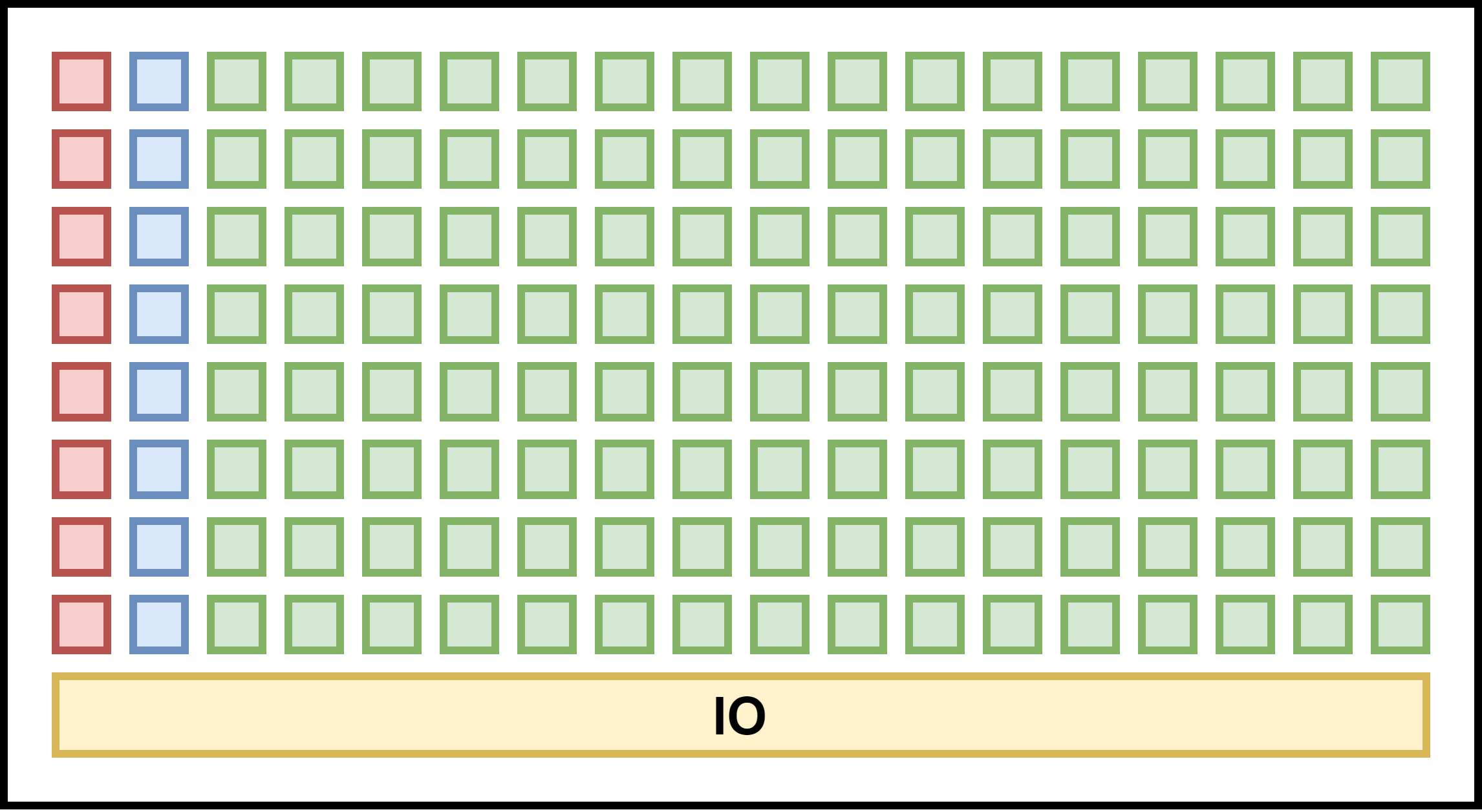}
    \caption{GPU Architecture}
\end{figure}

In the GPU diagram, the architecture emphasizes a much larger number of simple cores. Each core is optimized for performing specific, simple tasks in parallel, which is ideal for graphics rendering and other highly parallel computations. This design trades off individual core power for sheer numbers, focusing on throughput over latency.

\section{Conclusion}

In summary, CPUs and GPUs are both integral to modern computing, but they are optimized for different types of tasks. While CPUs are excellent for general-purpose, sequential operations, and multitasking, GPUs shine in parallel processing and large-scale data computations. Understanding the strengths of each will allow you to choose the right tool for the job, improving the efficiency and performance of your applications.
\chapter{Parallel Architectures Beyond GPUs}

\section{Understanding Parallelism in Computing}

Parallel computing is a type of computation where many calculations or processes are carried out simultaneously \cite{quinn1994parallel}.  Large problems, such as large language model(LLM)\cite{niu2024large,peng2024securinglargelanguagemodels,li2024surveyingmllmlandscapemetareview}, can often be divided into smaller ones, which can then be solved at the same time, leading to faster execution. In modern computing, parallelism is key to increasing performance. Parallel architectures are designed to efficiently execute these simultaneous tasks.

For example, when you open a web browser, different processes are happening simultaneously: loading images, fetching data from the server, and rendering the user interface. These tasks can be carried out in parallel, making the overall experience faster and smoother.

In Python, a basic example of parallelism can be seen using the \texttt{concurrent.futures} module, which allows you to run multiple tasks concurrently using threads or processes.

\begin{lstlisting}[style=python]
import concurrent.futures
import time

def task(n):
    print(f"Starting task {n}")
    time.sleep(1)
    print(f"Finished task {n}")

with concurrent.futures.ThreadPoolExecutor() as executor:
    tasks = [executor.submit(task, i) for i in range(5)]
\end{lstlisting}

In this example, 5 tasks run in parallel using threads. Without parallelism, these tasks would run one after the other, making the overall process slower.

\textcolor{codegreen}{Although the \texttt{concurrent.futures} module provides a mechanism for concurrency using multithreading, it's important to note that Python's Global Interpreter Lock (GIL) \cite{eggen2019thread} can limit true parallel execution for CPU-bound tasks. The GIL ensures that only one thread executes Python bytecode at a time, meaning that even with multiple threads, tasks may not be executed simultaneously. Therefore, multithreading is better suited for I/O-bound tasks, such as network requests or file operations, rather than CPU-bound tasks.}

For CPU-bound tasks that need to leverage multiple CPU cores, it is recommended to use \\ \texttt{concurrent.futures.ProcessPoolExecutor}, which bypasses the GIL by using multiple processes, enabling true parallelism. Here's an example using multiprocessing:

\begin{lstlisting}[style=python]
import concurrent.futures
import time

def task(n):
    print(f"Starting task {n}")
    time.sleep(1)
    print(f"Finished task {n}")

with concurrent.futures.ProcessPoolExecutor() as executor:
    tasks = [executor.submit(task, i) for i in range(5)]
\end{lstlisting}

In this example, the \texttt{ProcessPoolExecutor} creates multiple processes, with each task running in its process, thus achieving real parallel execution.

\section{Comparison of Parallel Architectures}

Different parallel architectures exist to handle tasks efficiently. Each of them is optimized for specific types of workloads. We will explore the most common architectures: GPUs, FPGAs, TPUs, and ASICs.

\subsection{GPU: Graphics and Beyond}

A GPU (Graphics Processing Unit) was initially designed to accelerate the rendering of images for video games and visual applications. However, because of its highly parallel structure, a GPU can also be used for general-purpose computation, especially tasks that can be divided into many smaller, independent tasks.

For instance, machine learning and scientific simulations take advantage of GPUs because these tasks require processing large amounts of data in parallel. Libraries like \texttt{TensorFlow} and \texttt{PyTorch} often use GPUs for accelerating neural network training.

\begin{lstlisting}[style=python]
import torch

# Check if GPU is available
device = "cuda" if torch.cuda.is_available() else "cpu"
print(f"Running on device: {device}")

# Simple tensor operation on GPU
tensor = torch.tensor([1.0, 2.0, 3.0]).to(device)
result = tensor * 2
print(result)
\end{lstlisting}

In this example, we perform a simple tensor operation on the GPU using PyTorch. The computation is offloaded to the GPU, leading to faster execution.

\subsection{FPGA: Customizable Hardware Parallelism}

An FPGA (Field-Programmable Gate Array) is a type of hardware that can be configured after manufacturing to perform specific tasks. FPGAs offer high parallelism by allowing custom architectures to be defined for specific algorithms, making them very efficient for specialized tasks like signal processing, cryptography, or real-time processing.

FPGAs are highly customizable, allowing developers to design parallel systems optimized for their exact needs, but they can be complex to work with due to the need for hardware-level programming languages like Verilog or VHDL.

The architecture of an FPGA is built around three main components: \textit{Configurable Logic Blocks (CLBs)} \cite{raj2020configurable}, \textit{Interconnects}, and \textit{Input/Output Blocks (IOBs)} \cite{swift2004dynamic}. CLBs are the fundamental units that can implement logic functions, and they are typically made up of Look-Up Tables (LUTs), flip-flops, and multiplexers. The LUTs are responsible for storing predefined logic functions. By programming the LUTs, an FPGA can implement any arbitrary logic function.

\textbf{How does an FPGA work?}

An FPGA operates by using its configurable logic blocks and interconnects, which can be reconfigured to implement any digital circuit. Here's a step-by-step breakdown of its working mechanism:

\begin{itemize}
    \item \textbf{Configurable Logic Blocks (CLBs):} CLBs are the core components of FPGAs that perform logic operations. Each CLB typically contains one or more LUTs, which act like small memories used to implement logic functions. For example, a LUT with four inputs can represent any 4-input Boolean function by storing the appropriate truth table values. In addition to LUTs, CLBs often include flip-flops for storing the state information and multiplexers for routing signals.
    
    \item \textbf{Programmable Interconnect:} The interconnect in an FPGA consists of a network of wires and programmable switches that connect the CLBs. By programming these interconnects, developers can define custom data paths between logic blocks, enabling the creation of complex circuits that are optimized for specific applications. The interconnects allow for parallel data flows, enabling FPGAs to exploit hardware-level parallelism.
    
    \item \textbf{Input/Output Blocks (IOBs):} The IOBs manage communication between the FPGA and external components such as sensors, processors, or other hardware. These blocks can be configured to support a variety of I/O standards, providing the flexibility needed for different applications.
    
    \item \textbf{Clocking and Timing:} FPGAs also feature a global clock network that synchronizes operations across the chip. Each CLB can be clocked at different rates, allowing for precise control over timing and data flow. Since timing is crucial in digital circuits, FPGAs often include dedicated clock management resources to ensure accurate synchronization.
\end{itemize}

This high degree of configurability allows FPGAs to excel in applications that demand parallelism and real-time processing. Because the hardware is programmable, multiple operations can occur simultaneously across different CLBs, resulting in significant performance improvements compared to traditional processors that execute instructions sequentially.

Below is an example of using Python to communicate with an FPGA for a custom hardware task.

\begin{lstlisting}[style=python]
# Example for controlling FPGA with Python using an interface like PySerial
import serial

# Connect to FPGA over serial communication
fpga_serial = serial.Serial('/dev/ttyUSB0', baudrate=115200, timeout=1)

# Send some data to FPGA
fpga_serial.write(b'Start Task\n')

# Read the response from FPGA
response = fpga_serial.readline().decode('utf-8')
print(f"FPGA response: {response}")
\end{lstlisting}

In this example, we use Python to send commands to an FPGA over a serial connection. The FPGA performs tasks as instructed and returns the result.

\subsection{TPU: Optimized for Machine Learning}

A TPU (Tensor Processing Unit) \cite{jouppi2017datacenter} is a specialized type of hardware developed by Google that is optimized for machine learning tasks, particularly those that use neural networks. TPUs are designed to accelerate the computation of tensor operations, which are at the heart of machine learning algorithms, such as matrix multiplications in deep learning. 

\textbf{How does a TPU work?}

At its core, a TPU is built to handle tensor operations very efficiently, which are the basic building blocks of neural network computations. These operations involve high-dimensional arrays of data (tensors) and are computationally intensive. A typical TPU architecture includes components like:

\begin{itemize}
    \item \textbf{Matrix Multiply Units (MXUs):} The TPU's most crucial component, the MXU, is designed to perform matrix multiplications much faster than traditional CPUs or GPUs. This is essential for deep learning models where matrix operations dominate the computation.
    
    \item \textbf{High Bandwidth Memory (HBM):} TPUs have high bandwidth memory that allows for faster access to large datasets, such as weights and activations in a neural network. This reduces the latency and increases the speed of computation during training and inference.
    
    \item \textbf{Software and Framework Integration:} TPUs are designed to work seamlessly with TensorFlow, Google’s machine learning framework. TensorFlow’s XLA \cite{chadha2017performance} (Accelerated Linear Algebra) compiler optimizes code for TPUs, ensuring that models take full advantage of the hardware’s capabilities.
\end{itemize}

TPUs are particularly effective when running large machine learning models, like those used in natural language processing, computer vision, or large-scale recommendation systems. Their architecture is specialized to handle repetitive tensor operations, making them faster and more efficient than general-purpose processors like CPUs or even GPUs for these specific tasks.

\textbf{Edge TPUs and Coral Processors}

In addition to cloud TPUs, Google has developed \textit{Edge TPUs}, which are designed for running machine learning models at the edge of the network, such as in IoT devices or mobile applications. These are specialized processors optimized for inference, rather than training, and have lower power consumption and cost compared to cloud TPUs. 

One example of an edge TPU is Google's \textbf{Coral TPU} \cite{sengupta2020high}, a hardware accelerator designed to run machine learning models directly on edge devices like smart cameras, sensors, or robotics. The Coral TPU can perform 4 trillion operations per second (TOPS) while using just a few watts of power, making it ideal for real-time applications in constrained environments.

\textbf{TPUs in NVIDIA GPUs}

NVIDIA, a leader in GPU (Graphics Processing Unit) technology, has also integrated features similar to TPUs into their GPUs, particularly for machine learning workloads. While traditional GPUs were originally designed for rendering graphics, NVIDIA’s recent generations of GPUs, such as those in the \textit{A100 Tensor Core} series, include specialized hardware blocks known as \textbf{Tensor Cores}. 

Tensor Cores are designed to accelerate the same types of matrix multiplications that TPUs focus on, particularly for deep learning tasks like convolutional neural networks (CNNs) and transformers. These cores allow NVIDIA GPUs to efficiently handle both training and inference tasks for machine learning models, providing competition to Google’s TPU technology in cloud and edge environments.

\textbf{Comparison between TPUs and GPUs}

While TPUs are custom-built for machine learning tasks, NVIDIA's GPUs with Tensor Cores offer more versatility, allowing them to be used in a wider variety of applications beyond just AI, including rendering, gaming, and scientific computing. However, for large-scale machine learning models, TPUs may offer better performance-per-watt and are specifically optimized for tensor operations, giving them an edge in highly specialized AI applications.

\textbf{Comparison between TPUs and GPUs}

While TPUs are custom-built for machine learning tasks, NVIDIA's GPUs with Tensor Cores offer more versatility, allowing them to be used in a wider variety of applications beyond just AI, including rendering, gaming, and scientific computing. However, for large-scale machine learning models, TPUs may offer better performance-per-watt and are specifically optimized for tensor operations, giving them an edge in highly specialized AI applications.

Regarding precision, TPUs often use reduced precision formats such as \textbf{bfloat16} (16-bit floating point) to increase computational efficiency and reduce power consumption, which generally results in faster computations. In contrast, GPUs, particularly those with Tensor Cores, support a wider range of precision formats, including \textbf{FP32}, \textbf{FP16}, and \textbf{INT8}, offering more flexibility.

While TPUs typically operate at lower precision, this does not usually lead to a significant drop in model performance, as many machine learning models can maintain accuracy with reduced precision. However, if higher precision is critical for a specific application, GPUs may have an advantage due to their broader precision support. \\

\textbf{Here is an example of using TPUs in Python with \texttt{TensorFlow}:}

\begin{lstlisting}[style=python]
import tensorflow as tf

# Check if TPU is available
resolver = tf.distribute.cluster_resolver.TPUClusterResolver.connect()
strategy = tf.distribute.TPUStrategy(resolver)

# Simple tensor operation on TPU
with strategy.scope():
    tensor = tf.constant([1.0, 2.0, 3.0])
    result = tensor * 2
    print(result)
\end{lstlisting}

In this example, we use TensorFlow to detect and use a TPU for running tensor computations.

This code is designed to run on a TPU and can be executed in Google Colab. To do so, ensure the TPU runtime is enabled by navigating to \textbf{Runtime} $\rightarrow$ \textbf{Change runtime type}, and selecting \textbf{TPU} as the hardware accelerator. TensorFlow will then detect the TPU and use it for tensor operations, improving computation speed. Be aware that some TensorFlow operations may have specific limitations or optimizations when run on a TPU.

\subsection{Other Architectures: ASICs and Beyond}

ASICs (Application-Specific Integrated Circuits) are chips designed for a specific task rather than general-purpose computation. Unlike FPGAs, ASICs cannot be reprogrammed after manufacturing. ASICs are extremely efficient for the tasks they are designed for, but they lack the flexibility of FPGAs and GPUs.

ASICs are often used in scenarios where performance and power efficiency are critical, such as in cryptocurrency mining or embedded systems for specific industrial applications.

For example, an ASIC (Application-Specific Integrated Circuit) designed for cryptocurrency mining can perform the specific calculations needed for Bitcoin mining far more efficiently than a general-purpose CPU or GPU.

\textbf{Advantages:} The primary advantage of ASICs is their speed and efficiency. Since they are purpose-built to perform the specific SHA-256 hashing operations \cite{gilbert2003security} used in Bitcoin mining, they can process these calculations much faster than CPUs or GPUs, which are general-purpose hardware designed to handle a wide range of computational tasks.

\textbf{Disadvantages:} Despite their speed, ASICs have significant drawbacks. One of the major disadvantages is resource waste. Bitcoin mining machines are frequently updated, and once a new, more efficient ASIC model is released, older slower, and less power efficient models become obsolete. Unlike GPUs, which can be repurposed for other computational tasks, ASICs are designed only to compute SHA-256 hashes. This lack of flexibility means that older ASICs quickly become useless and are essentially discarded as e-waste.

\section{Choosing the Right Architecture for Different Tasks}

When deciding which parallel architecture to use, consider the following factors:
\begin{itemize}
    \item \textbf{Task Type:} If your task involves simple but highly parallelizable computations, like matrix multiplication in deep learning, a GPU or TPU might be the best choice.
    \item \textbf{Custom Requirements:} If you need a highly specialized solution for tasks like signal processing or encryption, an FPGA or ASIC might be more suitable.
    \item \textbf{Development Complexity:} GPUs and TPUs are easier to program, as they are supported by high-level libraries like \texttt{TensorFlow} and \texttt{PyTorch}. FPGAs and ASICs require hardware-level programming, which is more complex and time-consuming.
    \item \textbf{Power Efficiency:} For tasks requiring low power consumption and high efficiency, ASICs are often the most efficient choice, while GPUs and FPGAs might consume more power.
\end{itemize}

In summary, understanding the nature of the task you are trying to parallelize is crucial in selecting the right architecture. Each architecture has its strengths and weaknesses, and the best choice depends on the specifics of your application.

\begin{itemize}
    \item Parallel Architectures
    \begin{itemize}
        \item GPU
        \begin{itemize}
            \item Machine Learning
            \item Graphics
        \end{itemize}
        \item FPGA
        \begin{itemize}
            \item Customizable Hardware
            \item Signal Processing
        \end{itemize}
        \item TPU
        \begin{itemize}
            \item Neural Networks
        \end{itemize}
        \item ASIC
        \begin{itemize}
            \item Cryptocurrency Mining
        \end{itemize}
    \end{itemize}
\end{itemize}
\chapter{Understanding Data Flow in Deep Learning: CPU, GPU, RAM, VRAM, Cache, and Disk Storage}

When training a deep learning model, the flow of data between different types of memory—such as CPU cache, GPU cache, CPU RAM (main memory), GPU VRAM (video memory), and disk storage—is critical to understand. Efficient data handling can significantly improve training performance. In this section, we will break down the flow of data step by step, including how caching mechanisms are involved.

\section{Understanding the Computer Memory Hierarchy}

Before diving into how data flows during the training of a deep learning model \cite{feng2024deeplearningmachinelearning,li2024deeplearningmachinelearning}, it's essential to understand the fundamental structure of a computer's memory system. Memory in a computer is organized hierarchically, often referred to as the "Memory Hierarchy Pyramid" \cite{jacob1996analytical}. This hierarchy ranges from the fastest but smallest memory types, like CPU caches, to the slowest but largest, such as disk storage. Each level of memory has distinct characteristics, and understanding these is critical for managing data efficiently during training.

\subsection{Memory Hierarchy Overview}

At the top of the pyramid, we have small but extremely fast memory, and as we go down the pyramid, memory becomes slower but larger in capacity. Below is an illustration of the typical memory hierarchy in modern computers:

\begin{tikzpicture}
% 金字塔
\fill[shade, top color=red, bottom color=green] (6.5,0) -- (8.5,4) -- (10.5,0) -- cycle;

% 右侧内容
\node[anchor=west] at (11.2,4) {CPU registers};
\node[anchor=west] at (11.2,3.2) {CPU Cache (L1, L2, L3)};
\node[anchor=west] at (11.2,2.4) {Main Memory (RAM)};
\node[anchor=west] at (11.2,1.6) {*GPU Memory (VRAM)};
\node[anchor=west] at (11.2,0.8) {Disk Storage (SSD/HDD)};
\node[anchor=west] at (11.2,0) {External Storage};

% 左侧添加成本和速度说明，使用 west 对齐
\node[anchor=west] at (3,4) {Cost: Very Expensive};
\node[anchor=west] at (3,3.2) {Cost: Expensive};
\node[anchor=west] at (3,2.4) {Cost: Medium};
\node[anchor=west] at (3,1.6) {Cost: Medium};
\node[anchor=west] at (3,0.8) {Cost: Cheap};
\node[anchor=west] at (3,0) {Cost: Very Cheap};

\node[anchor=west] at (0,4) {Speed: Fastest};
\node[anchor=west] at (0,3.2) {Speed: Fastest};
\node[anchor=west] at (0,2.4) {Speed: Fast};
\node[anchor=west] at (0,1.6) {Speed: Fast};
\node[anchor=west] at (0,0.8) {Speed: Slow};
\node[anchor=west] at (0,0) {Speed: Slowest};

\end{tikzpicture}

{\small \texttt{\textcolor{codegreen}{*GPU Memory (VRAM) shouldn't be put together with the CPU architecture, but for comparison, it's put together here.}}} \\

Let's break this down, step by step, starting from the top of the pyramid.

\subsection{CPU registers and Cache}

\textbf{What is it?} 

CPU registers and cache are the fastest and closest memory to the processor. Registers are small, high-speed storage locations directly inside the CPU, used to hold data that the processor is currently working on. They store temporary values such as instructions, operands, and intermediate results during computation. Because registers are too core CPU components and users will hardly ever touch them, we will not discuss registers here.

Cache \cite{smith1982cache}, on the other hand, is typically divided into different levels (L1, L2, and L3) that are progressively larger but slightly slower. The CPU uses the cache to store frequently accessed data or instructions to reduce the time it takes to access data from the slower main memory.

\textbf{Analogy:}  
Imagine you are cooking in your kitchen. The CPU cache is like the cooking utensils you keep right next to you on the countertop for easy access. You don’t need to walk to another room (main memory) to grab them because they are right where you need them most frequently.

\textbf{Real-world Example:}  
Suppose you have a loop in Python that performs repeated calculations. If the data being processed fits within the CPU cache, it will run much faster because the CPU can repeatedly access this data without fetching it from the slower RAM.

\begin{lstlisting}[style=python]
# Example: Repeated Access to Cached Data
data = [i for i in range(1000)]  # Assuming this fits in cache
result = 0
for value in data:
    result += value * 2  # Fast repeated access
\end{lstlisting}

\textbf{Key Points:}
\begin{itemize}
    \item Extremely fast (nanosecond access times).
    \item Very small (usually a few MBs).
    \item Frequently accessed data is stored here to avoid accessing slower memory.
\end{itemize}

\subsection{Main Memory (RAM)}

\textbf{What is it?}  
RAM (Random Access Memory) \cite{memory2013ram}   is the computer's primary working memory. It is much larger than the CPU cache, but it is slower in comparison. When a program runs, its data and instructions are loaded into RAM from disk storage, allowing faster access by the CPU.

\textbf{Analogy:}  
Continuing with the kitchen analogy, RAM is like the ingredients you keep in your kitchen pantry. You don’t need to go to the grocery store (disk storage) every time you need something because it's already close by, but it still takes more time to retrieve than grabbing something from the countertop (CPU cache).

\textbf{Real-world Example:}  
When you load a dataset into memory for training a deep learning model, the data is usually stored in RAM. The model will fetch batches of data from RAM, and this is considerably faster than fetching data directly from disk.

\begin{lstlisting}[style=python]
# Example: Loading data into RAM
import numpy as np

data = np.random.random((10000, 100))  # Load data into RAM
for i in range(100):
    batch = data[i*100:(i+1)*100, :]   # Fetch batches from RAM
    # Process batch for model training
\end{lstlisting}

\textbf{Key Points:}
\begin{itemize}
    \item Fast but slower than cache (access time in microseconds).
    \item Larger than cache (several GBs to TBs).
    \item Used to store data currently in use by running programs.
\end{itemize}

\subsection{GPU Memory (VRAM)}

\textbf{What is it?}  
Graphics Processing Units (GPUs) are often used to accelerate deep learning computations, and they have their dedicated memory, called Video RAM (VRAM) \cite{jones1992memory}. VRAM is faster than RAM because it is located closer to the GPU, which can be accessed it without going through the CPU.

\textbf{Analogy:}  
VRAM is like having a special section in your kitchen just for your assistant chef (the GPU). They can access ingredients faster than you can because their section is optimized for what they need, just as the GPU’s memory is optimized for parallel processing tasks.

\textbf{Real-world Example:}  
When training a deep learning model on a GPU, the model and the data must be loaded into the GPU’s VRAM. If the dataset is too large for the VRAM, the model must offload some data back to RAM or disk, which slows down training.

\begin{lstlisting}[style=python]
# Example: Loading model and data onto GPU (VRAM)
import torch

device = torch.device('cuda' if torch.cuda.is_available() else 'cpu')
data = torch.randn(10000, 100).to(device)  # Load data into GPU VRAM
model = torch.nn.Linear(100, 10).to(device)  # Load model into GPU VRAM

output = model(data)  # Process data using the model on the GPU
\end{lstlisting}

\textbf{Key Points:}
\begin{itemize}
    \item Very fast (nanoseconds).
    \item Dedicated for GPU operations.
    \item Typically smaller than RAM (a few GBs to 24 GB in consumer GPUs).
\end{itemize}

\subsection{Differences between Main Memory (RAM) and GPU Memory (VRAM)}

Main memory (RAM) and graphics memory (VRAM) play different roles in modern computing systems. While both are types of volatile memory used for temporary data storage, they have significant differences in architecture, performance, and usage.

First, the \textbf{cost and replaceability} are key differences between the two. RAM is generally cheaper than VRAM and can often be replaced or upgraded by the user to improve system performance. VRAM, on the other hand, is tightly integrated with the GPU and typically cannot be upgraded separately; to increase VRAM capacity, the entire graphics card must be replaced.

\textbf{Interaction with CPU and GPU} is another distinguishing factor. RAM communicates directly with the central processing unit (CPU), storing program data and temporary information needed for various computational tasks. In contrast, VRAM is specifically designed for the graphics processing unit (GPU) to handle image and video rendering tasks. Due to the high parallelism and large data requirements of GPUs, VRAM is designed with higher bandwidth and lower latency to quickly access and process graphical data.

The \textbf{complexity of data transfer} between the two is also an important distinction. If data needs to be transferred from RAM to VRAM, it must typically pass through the CPU. The CPU coordinates the process of fetching data from the main memory and transmitting it to the GPU’s VRAM over the PCIe bus. While this process is relatively efficient, it still introduces some latency due to the transfer.

There is also a difference in \textbf{physical location and bandwidth}. RAM is typically located on the CPU side and connects to the CPU through memory buses like DDR. VRAM, on the other hand, surrounds the GPU. Because of this close integration, VRAM typically has higher bandwidth than RAM to meet the demands of large data transfers during GPU-intensive tasks. For example, GDDR (Graphics DDR) \cite{kim2007performance} memory often has significantly higher bandwidth than standard DDR memory. This high bandwidth allows the GPU to handle complex graphical rendering and video processing without becoming bottlenecked by slow data transfer rates.

In conclusion, RAM and VRAM serve distinct purposes, each optimized for the performance needs of the CPU and GPU, respectively.

\subsection{Disk Storage (SSD/HDD)}

\textbf{What is it?}  
Disk storage, whether Solid-State Drives (SSD) or Hard Disk Drives (HDD), is where all data and programs are permanently stored. While SSDs are faster than HDDs, both are still much slower than RAM and are generally only used to store data that is not actively being processed.

\textbf{Analogy:}  
Disk storage is like your refrigerator or the grocery store. You store ingredients that you don't immediately need in the refrigerator, and when you run out of something, you go to the grocery store to restock. However, fetching items from the refrigerator (SSD) is faster than making a trip to the grocery store (HDD).

\textbf{Real-world Example:}  
When working with very large datasets that don’t fit into RAM, deep learning workflows may load data in smaller chunks from disk storage. This process is slower but necessary for large-scale tasks.

\begin{lstlisting}[style=python]
# Example: Loading data from disk in chunks
import pandas as pd

for chunk in pd.read_csv('large_dataset.csv', chunksize=1000):
    # Process each chunk in memory
    process(chunk)
\end{lstlisting}

\textbf{Key Points:}
\begin{itemize}
    \item Very large (up to several TBs).
    \item Slow (milliseconds).
    \item Used for permanent storage of data and programs.
\end{itemize}

\subsection{External Storage}

\textbf{What is it?}  
External storage includes devices like USB drives, external hard drives, or cloud storage. These are typically used for backups or transferring data between systems and are not directly accessed during normal program execution.

\textbf{Analogy:}  
External storage is like a warehouse where you store things you don’t need very often. It takes considerable time to retrieve something from the warehouse, so you only do so when necessary.

\textbf{Real-world Example:}  
When collaborating with other teams, you might store your trained model or datasets in cloud storage. Accessing this data takes time, and it is typically copied to a local disk or memory for faster access.

\textbf{Key Points:}
\begin{itemize}
    \item Very large (TBs to PBs in cloud storage).
    \item Very slow (due to network latency).
    \item Used for backups and data transfer.
\end{itemize}

\subsection{Conclusion: The Importance of Memory Hierarchy in Deep Learning}

Understanding the computer memory hierarchy is essential when working with deep learning models. Efficient data management across different types of memory can significantly impact the speed and performance of training. For instance, keeping as much data as possible in the faster levels of the hierarchy (e.g., CPU cache, RAM, or VRAM) will reduce the time spent fetching data from slower storage layers (e.g., disk storage). 

By optimizing the way we load and process data, we can make better use of the available memory resources and reduce the time it takes to train deep learning models.

\section{Data Storage on Disk}

Initially, the data you are using for training, such as images or text files, is typically stored on your system's hard drive or SSD. Hard drives (especially HDDs) are the slowest form of memory because they rely on mechanical parts or flash memory. Though SSDs are faster, they still lag behind RAM or GPU memory in terms of speed.

For instance, consider a dataset of 10,000 images stored on your disk.

\begin{lstlisting}[style=python]
import os
data_path = "/path/to/dataset"
images = os.listdir(data_path)
\end{lstlisting}

At this point, the images are just filenames sitting on the disk. To work with them, you need to load these files into RAM.

\section{Loading Data into RAM (CPU Memory)}

The next step is to load the data from the disk into CPU RAM. RAM (Random Access Memory) is much faster than disk storage and allows the CPU to access and process the data more efficiently.

\begin{lstlisting}[style=python]
from PIL import Image

# Load an image into RAM
image_path = os.path.join(data_path, images[0])
image = Image.open(image_path)
\end{lstlisting}

Once loaded, the image resides in the CPU's RAM, ready for further processing. However, the CPU may not always access the RAM directly for every single operation due to its internal cache system.

\section{CPU Cache: Faster Memory Access}

The CPU has several levels of cache (L1, L2, and sometimes L3), which are much smaller and faster than RAM. The CPU cache is used to temporarily store data that is frequently accessed or recently used, speeding up operations by reducing the need to constantly access slower RAM.

When your deep learning framework, such as PyTorch or TensorFlow, loads small batches of data, the CPU might place part of that data in its cache to accelerate further computation.

\begin{tikzpicture}
\draw[->, thick] (0,0) -- (2,0) node[midway, above] {Disk};
\draw[->, thick] (2,0) -- (4,0) node[midway, above] {RAM};
\draw[->, thick] (4,0) -- (6,0) node[midway, above] {CPU Cache};
\end{tikzpicture}

This way, the CPU only has to load data from the RAM once, and subsequent operations are done using the cache, which is much faster.

\subsection{L3 Cache Allocation Strategies and Their Impact}

L3 cache differs from L1 and L2 caches in that it is often shared among multiple CPU cores, making it a critical resource in modern multi-core processors. The way the L3 cache is allocated across these cores can have a significant effect on both CPU and GPU-accelerated tasks.

Two main strategies for L3 cache allocation are:

\begin{formatteditem}
    \item \textbf{Static Allocation}: This method assigns a fixed portion of the L3 cache to each core or process. While predictable, static allocation can lead to inefficient cache usage if some cores are underutilized, leaving portions of the cache unused.
    \item \textbf{Dynamic Allocation}: In this strategy, the L3 cache is distributed based on real-time needs. Cores or processes that require more cache can dynamically receive a larger share, optimizing for overall system performance by reducing cache misses.
\end{formatteditem}

\subsubsection{Impact on Deep Learning and GPGPU Workloads}

In deep learning frameworks like PyTorch or TensorFlow, dynamic L3 cache allocation can be particularly beneficial when pre-processing data or managing multiple threads. For example, when CPU threads handle data preprocessing before transferring it to the GPU, dynamic cache allocation allows more cache space to be allocated to the most active threads, reducing memory latency and speeding up data transfers.

For GPGPU tasks, managing cache contention between multiple cores becomes important. If a large dataset is being transferred to the GPU, dynamic allocation can reduce bottlenecks by optimizing cache usage across the CPU cores handling different parts of the data.

On the other hand, static cache allocation might be preferable for real-time inference tasks, where maintaining predictable latency is crucial. By assigning a fixed portion of the cache to each core, the system can ensure consistent performance, which is critical in applications like autonomous driving or real-time video analysis.

Choosing between static and dynamic L3 cache allocation depends on the workload. Dynamic allocation often improves performance in data-intensive, parallel tasks, especially when working with deep learning and GPGPU workloads. However, static allocation may be more suitable for tasks requiring predictable performance, such as real-time inference. By optimizing L3 cache usage, overall system efficiency can be improved, leading to faster data processing and reduced latency in both CPU and GPU operations.

\section{Transferring Data to the GPU (VRAM and GPU Cache)}

Once the data is in RAM and possibly cached by the CPU, the next step is to send it to the GPU for computation. The GPU has its memory, called VRAM (Video Random Access Memory), which is optimized for parallel tasks such as deep learning matrix computations.

In addition to VRAM, the GPU also has its cache hierarchy (L1, L2 cache). When processing large datasets, frequently accessed data may be stored in the GPU’s cache, much like in the CPU.

\begin{lstlisting}[style=python]
import torch

# Convert the image to a tensor and send it to GPU
image_tensor = torch.tensor(image).to('cuda')
\end{lstlisting}

Here, the image tensor is moved from the CPU to the GPU’s VRAM. Once in the VRAM, the data can be cached in the GPU’s L1 or L2 cache, allowing the GPU to perform computations much faster by minimizing access to VRAM.

\section{Data Flow during Training}

During training, batches of data are repeatedly loaded from the disk into RAM and then moved to VRAM for GPU computation. The GPU uses both its VRAM and cache to efficiently process the data, perform computations, and update model parameters. After each computation, results (such as loss values or updated model weights) may be sent back to the CPU for logging or further analysis.

Here is a typical data flow during training:

\begin{tikzpicture}
\draw[->, thick] (0,0) -- (3,0) node[midway, above] {Disk to RAM};
\draw[->, thick] (3,0) -- (7,0) node[midway, above] {RAM to CPU Cache};
\draw[->, thick] (7,0) -- (11,0) node[midway, above] {CPU Cache to VRAM};
\draw[->, thick] (11,0) -- (15,0) node[midway, above] {VRAM to GPU Cache};
\end{tikzpicture}

This flow continues in each batch of data. If data is used repeatedly (e.g., in multiple epochs), it will likely remain in the caches, further speeding up subsequent operations.

\section{Example Workflow: Training a Simple Neural Network}

Let’s walk through an example in PyTorch to see how the data moves through different memory locations.

\begin{lstlisting}[style=python]
import torch
from torch.utils.data import DataLoader
from torchvision import datasets, transforms

# Loading the dataset into RAM
transform = transforms.Compose([transforms.ToTensor()])
train_data = datasets.MNIST(root='./data', train=True, download=True, transform=transform)
train_loader = DataLoader(train_data, batch_size=32, shuffle=True)

# Model and optimizer setup (model will be transferred to GPU)
model = torch.nn.Linear(28*28, 10).to('cuda')
optimizer = torch.optim.SGD(model.parameters(), lr=0.01)

# Training loop
for epoch in range(5):
    for batch in train_loader:
        inputs, labels = batch

        # Data from RAM to CPU cache
        inputs, labels = inputs.view(inputs.size(0), -1), labels

        # Data from CPU to GPU (VRAM)
        inputs = inputs.to('cuda')
        labels = labels.to('cuda')

        # Forward pass and backpropagation
        optimizer.zero_grad()
        outputs = model(inputs)
        loss = torch.nn.functional.cross_entropy(outputs, labels)
        loss.backward()

        # Optimization step (done on GPU)
        optimizer.step()
\end{lstlisting}

In this example:

1. The MNIST dataset is loaded from the disk to RAM.
2. Small batches of data are likely cached in the CPU cache during training to speed up operations.
3. Data is moved from the CPU (RAM or cache) to the GPU's VRAM for processing.
4. The GPU caches frequently accessed data to avoid unnecessary reads from VRAM.
5. After processing, the results are sent back to the CPU for logging and analysis.

\section{Conclusion}

Understanding the flow of data across CPU cache, GPU cache, RAM, VRAM, and disk storage is critical for optimizing deep learning models. Here is a summary of the memory components involved in this process:

\begin{itemize}
    \item \textbf{Disk Storage}: Slowest; where the data is initially stored.
    \item \textbf{RAM (CPU Memory)}: Faster than disk, used to hold data before it is sent to the CPU or GPU.
    \item \textbf{CPU Cache (L1, L2)}: Very fast; used to store frequently accessed data to reduce latency.
    \item \textbf{VRAM (GPU Memory)}: Used to hold data being processed by the GPU.
    \item \textbf{GPU Cache (L1, L2)}: Fastest; used by the GPU for quick access to frequently used data.
\end{itemize}

By understanding how data flows through these memory locations, you can optimize your training process, reduce bottlenecks, and fully utilize your hardware resources.

\chapter{Deep Dive into GPU Architecture}

\section{The GPU Hierarchical Structure}

This section will explain the fundamental structure of modern GPUs and introduce key components such as the core and processing units. Unlike CPUs, which are optimized for serial tasks, GPUs excel at handling large-scale parallelism, making them suitable for data-intensive operations such as image processing, machine learning \cite{Peng_2024}, and scientific simulations.

A GPU consists of several key components:

\begin{itemize}
    \item \textbf{Cores}: The fundamental processing units capable of executing individual instructions.
    \item \textbf{Streaming Multiprocessors (SMs)}: Groups of cores that work together to execute tasks in parallel.
    \item \textbf{Memory Hierarchy}: Different levels of memory, such as global memory, shared memory, and registers, optimize data access during computation.
\end{itemize}

\subsection{Overview of GPU Processing Pipeline}

The GPU processing pipeline consists of multiple stages, each performing a specific task in the processing of data. From the moment data enters the GPU until it produces the final output, the pipeline ensures efficient parallel execution.

Here is a simplified overview of the GPU processing pipeline:

\begin{enumerate}
    \item \textbf{Data Input}: Data is passed to the GPU, typically from the system's main memory (RAM) to the GPU's global memory.
    \item \textbf{Kernel Execution}: The GPU processes data through kernels, which are small programs that run in parallel across thousands of cores.
    \item \textbf{Thread Management}: Threads are assigned to the cores, where each thread processes a small portion of the overall data.
    \item \textbf{Memory Access}: Data is read from and written to the memory hierarchy, including shared memory and registers for optimal performance.
    \item \textbf{Data Output}: After processing, the data is returned to the system memory or used in further GPU computations.
\end{enumerate}

The GPU processing pipeline differs from the CPU pipeline mainly in how it handles parallelism. While a CPU focuses on executing a few instructions quickly with a deep pipeline and high clock speeds, the GPU prioritizes executing many instructions simultaneously with many cores working in parallel.

\subsection{Streaming Multiprocessors (SMs)}

At the heart of modern GPUs are Streaming Multiprocessors (SMs). Each SM contains a group of cores that can execute threads in parallel, allowing the GPU to handle massive parallelism efficiently. An SM consists of:

\begin{itemize}
    \item \textbf{Cores}: Individual processing units that execute threads.
    \item \textbf{Warp Scheduler}: Determines how threads are grouped into warps, which are sets of 32 threads that execute the same instruction simultaneously.
    \item \textbf{Registers}: Fast, low-latency memory used by individual threads for storing variables.
    \item \textbf{Shared Memory}: A small, user-managed cache that allows threads within a block to share data and communicate efficiently.
\end{itemize}

Each SM is designed to execute multiple threads simultaneously, maximizing parallelism and ensuring that the GPU can handle tasks like matrix multiplications, convolution operations, and other compute-heavy tasks in parallel.

\section{Understanding Grid and Blocks in CUDA}

In CUDA (Compute Unified Device Architecture), computation is organized into a hierarchy of grids and blocks. This hierarchical structure allows the GPU to break down complex problems into smaller, manageable pieces, enabling large-scale parallelism.

\subsection{Defining the Grid}

The \textbf{grid} is the highest-level abstraction in CUDA's execution model. It represents the overall problem space that is being solved on the GPU. A grid is composed of multiple \textbf{blocks}, which are further subdivided into \textbf{threads}.

For example, if you are performing an image processing task on a 1024x1024 pixel image, the entire image could be represented as a grid, where each pixel is processed by a different thread.

\subsection{Blocks: The Subdivision of Grids}

Each grid in CUDA is subdivided into \textbf{blocks}. A block is a collection of threads that can execute independently on the GPU. Each block is assigned to an SM, which executes the threads of that block in parallel. 

A key feature of blocks is that they can communicate with each other using \textbf{shared memory}, which is faster than accessing global memory. This allows blocks to collaborate on a subset of the overall computation.

For example, suppose you are performing matrix multiplication. The grid could represent the overall matrix, and each block could handle the multiplication of a submatrix, with threads working on individual elements within that submatrix.

Here's a Python code example illustrating how to define a grid and blocks in CUDA:

\begin{lstlisting}[style=python]
import numpy as np
from numba import cuda

# Define the kernel function to run on the GPU
@cuda.jit
def matrix_addition(a, b, result):
    # Get the index of the current thread in the grid
    i, j = cuda.grid(2)
    
    # Perform addition if the index is within the bounds
    if i < result.shape[0] and j < result.shape[1]:
        result[i, j] = a[i, j] + b[i, j]

# Initialize data
N = 1024
a = np.random.rand(N, N)
b = np.random.rand(N, N)
result = np.zeros((N, N))

# Define the grid and block dimensions
threads_per_block = (16, 16)  # A block is 16x16 threads
blocks_per_grid_x = (a.shape[0] + threads_per_block[0] - 1) // threads_per_block[0]
blocks_per_grid_y = (a.shape[1] + threads_per_block[1] - 1) // threads_per_block[1]
blocks_per_grid = (blocks_per_grid_x, blocks_per_grid_y)

# Launch the kernel
matrix_addition[blocks_per_grid, threads_per_block](a, b, result)

print(result)
\end{lstlisting}

In this example, a 1024x1024 matrix addition is performed on the GPU. The grid is defined as a collection of 64x64 blocks, and each block contains 16x16 threads, each responsible for a small portion of the matrix.

\section{Threads and Warps}

In this section, we will explore how threads are organized and executed on a GPU using CUDA. We will also discuss the key concepts behind their management for achieving optimal computational performance.

\subsection{What is a Thread?}

A thread is the smallest unit of computation in CUDA, the parallel computing platform from NVIDIA. Each thread runs a specific portion of code, typically working on one element of data. While a single thread can execute a small task, CUDA is designed to handle thousands of threads simultaneously, allowing it to solve complex problems by breaking them down into smaller tasks.

For example, let's say we want to add two arrays together, element by element. Each thread will handle the addition of one element from each array. Here's how we might set this up in CUDA:

\begin{lstlisting}[style=cpp, caption={C/C++ Code With CUDA}]
__global__ void add_arrays(float *a, float *b, float *result) {
    int index = threadIdx.x;
    result[index] = a[index] + b[index];
}
\end{lstlisting}

In this code:
\begin{itemize}
    \item Each thread works on a single element in the arrays \texttt{a} and \texttt{b}.
    \item \texttt{threadIdx.x} gives each thread a unique index within its block, allowing it to access and process different elements in the array.
\end{itemize}

CUDA uses this kind of thread-based parallelism to accelerate computations, making it possible to perform complex operations much faster than a single CPU could.

\subsection{Warps: Groups of Threads}

In CUDA, threads are grouped in units called \textbf{warps}. A warp consists of 32 threads, which are executed simultaneously by the GPU's scheduler. Warps are crucial because the GPU processes threads in groups, and efficient warp-level execution leads to optimal performance.

When a warp is scheduled, all 32 threads within it are executed in lockstep, meaning they perform the same instruction at the same time. However, each thread can operate on different data. If all threads in a warp execute the same code path without branching, the GPU can achieve maximum efficiency.

For instance, in the example above, if all threads are simply adding elements of two arrays, the GPU can schedule the warp to execute the additions simultaneously for all threads.

If threads within a warp start taking different branches (i.e., different execution paths), it can lead to what is called thread divergence, which we will discuss next.

\subsection{Managing Thread Divergence}

Thread divergence occurs when threads within a warp follow different execution paths. For example, if some threads in a warp take one branch of an \texttt{if} statement, while others take a different branch, the warp must execute both branches sequentially, reducing the GPU’s efficiency.

Consider the following code:

\begin{lstlisting}[style=cpp, caption={C/C++ Code With CUDA}]
__global__ void compute(float *data, int *flags) {
    int index = threadIdx.x;
    
    if (flags[index] == 1) {
        data[index] = data[index] * 2;
    } else {
        data[index] = data[index] + 1;
    }
}
\end{lstlisting}

If some threads within the warp evaluate \texttt{flags[index] == 1} as true while others evaluate it as false, the warp will need to execute both paths, which leads to thread divergence.

\textbf{Strategies to minimize divergence:}
- Avoid conditional branching whenever possible, especially within a warp.
- Try to structure your code so that threads in a warp follow the same execution path.
- When branching is unavoidable, try to group threads with similar execution paths into the same warp.

By minimizing thread divergence, you can ensure more efficient execution of warps and better overall performance on the GPU.

\section{Memory Hierarchy in GPUs}

GPUs have a complex memory hierarchy, designed to balance between capacity, latency, and access speed. Understanding this hierarchy is key to optimizing performance when writing CUDA code. In this section, we will look at the different levels of memory, including global memory, shared memory, and registers.

\subsection{Global Memory}

\textbf{Global memory} is the main memory space on a GPU and has a large capacity. All threads can access global memory, but it has relatively high latency, meaning it takes longer for data to be retrieved from it compared to other types of memory. 

Global memory is typically used for storing large amounts of data, such as arrays or matrices that need to be shared across many threads.

\begin{lstlisting}[style=cpp, caption={C/C++ Code With CUDA}]
__global__ void scale_array(float *arr, float factor) {
    int index = threadIdx.x;
    arr[index] *= factor;  // Accessing global memory
}
\end{lstlisting}

In this example, the array \texttt{arr} is stored in global memory, and all threads can access and modify it. To make your code more efficient, try to minimize the number of accesses to global memory, as each access incurs a high latency cost.

\subsection{Shared Memory}

\textbf{Shared memory} is a fast, low-latency memory space that is shared among all threads within the same block. Unlike global memory, shared memory is much faster to access, and can be used for temporary storage during computations. 

Using shared memory allows threads within the same block to collaborate and exchange data efficiently.

\begin{lstlisting}[style=cpp, caption={C/C++ Code With CUDA}]
__global__ void compute_sum(float *input, float *output) {
    __shared__ float temp[BLOCK_SIZE];
    int index = threadIdx.x;

    temp[index] = input[index];
    __syncthreads();  // Ensure all threads have written to shared memory

    if (index == 0) {
        float sum = 0;
        for (int i = 0; i < BLOCK_SIZE; i++) {
            sum += temp[i];
        }
        output[0] = sum;
    }
}
\end{lstlisting}

In this example, the array \texttt{temp} is stored in shared memory, allowing the threads in the block to quickly read and write intermediate results. This technique reduces the need for global memory access, significantly speeding up the computation.

\subsection{Registers and Local Memory}

Each thread has access to a small number of \textbf{registers}, which are the fastest type of memory on the GPU. Registers store variables used by a single thread, providing very fast access. However, the number of registers is limited, and when a thread needs more memory than is available in its registers, it uses \textbf{local memory}.

Local memory is slower than registers, but still faster than global memory. It is used to store temporary variables that are too large to fit in registers.

\begin{lstlisting}[style=cpp, caption={C/C++ Code With CUDA}]
__global__ void process_data(float *input, float *output) {
    float temp = input[threadIdx.x];  // Stored in a register
    output[threadIdx.x] = temp * 2;
}
\end{lstlisting}

In this example, the variable \texttt{temp} is stored in a register, allowing fast access for computations. Registers are ideal for storing frequently used variables, and they help speed up the execution of each thread.

In summary, understanding how to efficiently use global memory, shared memory, and registers is critical for writing high-performance CUDA code.
\section{Hierarchy of Grid, Block, and Thread in GPUs}

In GPU architectures, threads are organized into a hierarchy for efficient parallel execution. The highest level of organization is the \textbf{Grid}, which consists of multiple \textbf{Blocks}. Each Block contains multiple \textbf{Threads}, the smallest units of execution. The diagram below illustrates this hierarchy:

\begin{itemize}
    \item \textbf{Grid}: A collection of blocks, allowing large-scale parallel computations by distributing tasks across multiple blocks.
    \item \textbf{Block}: A group of threads that can share data via shared memory and be synchronized within the block. This example includes 16 blocks.
    \item \textbf{Thread}: The smallest unit of execution. Each block contains 9 threads (3x3 structure in this example).
\end{itemize}

\begin{center}
\begin{tikzpicture}[
    thread/.style={rectangle, draw, minimum size=0.5cm, fill=blue!20},
    block/.style={rectangle, draw, minimum size=2.0cm, rounded corners, fill=green!20, thick},
    grid/.style={rectangle, draw, thick, rounded corners, fill=red!20},
    node distance=0.6cm
]

% Grid
\node[grid, minimum width=10.5cm, minimum height=10.5cm, label={above:Grid}] (grid) {};

% Generate 16 blocks in a 4x4 grid, inside the main grid
\foreach \i in {0,1,2,3}
    \foreach \j in {0,1,2,3}
        \node[block] at ([shift={(\i*2.5+1.5,-\j*2.5-1.5)}]grid.north west) (block\i\j) {};

% Threads within each Block (for block at 0,0 as an example)
\foreach \x in {0.5, 1.0, 1.5}
    \foreach \y in {0.5, 1.0, 1.5}
        \node[thread] at ([shift={(\x,-\y)}]block00.north west) {};

% Threads in the other blocks (similar to block00)
\foreach \i in {0,1,2,3}
    \foreach \j in {0,1,2,3}
        \foreach \x in {0.5, 1.0, 1.5}
            \foreach \y in {0.5, 1.0, 1.5}
                \node[thread] at ([shift={(\x,-\y)}]block\i\j.north west) {};

\end{tikzpicture}
\end{center}

\section{Extended Hierarchy of Cluster, GPUs, and SMs in Blackwell Architecture}

In advanced GPU architectures like NVIDIA's Blackwell, the structure is more intricate, with multiple levels of hierarchy extending from clusters to GPUs and SMs. Each cluster consists of several \textbf{GPUs}, and each GPU is organized into multiple \textbf{Streaming Multiprocessors (SMs)}, which handle the execution of threads in parallel. This hierarchy can be extended further to represent network interconnections within the system.

\begin{itemize}
    \item \textbf{Cluster}: Represents the highest level of hierarchy, consisting of multiple GPUs interconnected through a high-speed network.
    \item \textbf{GPU}: Each GPU contains several SMs. In this example, we assume there are 4 GPUs per cluster.
    \item \textbf{SM (Streaming Multiprocessor)}: Each GPU contains multiple SMs, which can execute threads in parallel. Each SM typically contains several cores and scheduling units.
\end{itemize}

\begin{center}
\begin{tikzpicture}[
    thread/.style={rectangle, draw, minimum size=0.5cm, fill=blue!20},
    block/.style={rectangle, draw, minimum size=2.0cm, rounded corners, fill=green!20, thick},
    grid/.style={rectangle, draw, thick, rounded corners, fill=red!20},
    node distance=0.6cm
]

% Grid
\node[grid, minimum width=10.5cm, minimum height=10.5cm, label={above:Cluster}] (grid) {};

% Generate 16 blocks in a 4x4 grid, inside the main grid
\foreach \i in {0,1,2,3}
    \foreach \j in {0,1,2,3}
        \node[block] at ([shift={(\i*2.5+1.5,-\j*2.5-1.5)}]grid.north west) (block\i\j) {};

% Threads within each Block (for a block at 0,0 as an example)
\foreach \x in {0.5, 1.0, 1.5}
    \foreach \y in {0.5, 1.0, 1.5}
        \node[thread] at ([shift={(\x,-\y)}]block00.north west) {};

% Threads in the other blocks (similar to block00)
\foreach \i in {0,1,2,3}
    \foreach \j in {0,1,2,3}
        \foreach \x in {0.5, 1.0, 1.5}
            \foreach \y in {0.5, 1.0, 1.5}
                \node[thread] at ([shift={(\x,-\y)}]block\i\j.north west) {};

\end{tikzpicture}
\end{center}

\chapter{GPU Algorithms and Parallel Programming}

\section{Introduction to Parallel Programming in CUDA}

This section provides an overview of parallel programming concepts and how they are implemented using NVIDIA's CUDA architecture for efficient computation on GPUs.

\subsection{What is Parallel Programming?}
Parallel programming is a programming paradigm where many processes are executed simultaneously. It is widely used to speed up computational tasks by dividing work across multiple processors. Modern GPUs (Graphics Processing Units) are designed for such tasks, allowing thousands of threads to execute concurrently, making them ideal for scientific computing, machine learning, image processing, and more.

CUDA (Compute Unified Device Architecture) is NVIDIA’s parallel computing architecture that allows developers to use the GPU for general-purpose processing. In CUDA programming, the term "kernel" refers to a function that runs on the GPU, which is executed by many threads in parallel. 

\subsection{How CUDA Works}

At the core of CUDA programming are the following components:

\begin{itemize}
\item \textbf{Host (CPU)}: This refers to the system's central processing unit (CPU). The CPU typically launches the parallel tasks, manages memory, and interacts with the GPU.
\item \textbf{Device (GPU)}: The GPU is the device where kernels (CUDA functions) are executed.
\item \textbf{Kernel}: A function written to be executed on the GPU. The kernel is launched by the CPU and runs on multiple threads in parallel.
\item \textbf{Threads and Blocks}: A kernel is executed by many lightweight threads. Threads are grouped into blocks, and blocks are organized into grids. The GPU schedules and runs these threads concurrently to maximize performance.
\end{itemize}

\subsection{Writing Your First CUDA Program}

Let’s begin by writing a simple CUDA program that adds two arrays in parallel. The example demonstrates how to write, compile, and execute a basic CUDA program. 

Below is the structure of a simple CUDA program:

\begin{lstlisting}[style=cpp, caption={C/C++ Code With CUDA}]
// A simple CUDA kernel to add two arrays
__global__ void add(int *a, int *b, int *c) {
    int index = threadIdx.x;  // Get the thread ID
    c[index] = a[index] + b[index];  // Perform addition
}
\end{lstlisting}

\textbf{Explanation:}

\begin{itemize}
\item \_\_global\_\_: This indicates that the function (kernel) is meant to be run on the GPU.
\item threadIdx.x: Each thread has a unique ID, and here we are using that ID to identify which elements of the arrays to add.
\item The kernel operates on arrays element by element in parallel.
\end{itemize}

The kernel is just a part of the program. We also need to allocate memory for arrays on the GPU, copy data from the CPU to the GPU, launch the kernel, and finally copy the result back to the CPU.

Here is the complete program:

\begin{lstlisting}[style=cpp, caption={C/C++ Code With CUDA}]
#include <stdio.h>

// Kernel function to add the elements of two arrays
__global__ void add(int *a, int *b, int *c) {
    int index = threadIdx.x;
    c[index] = a[index] + b[index];
}

int main() {
    int a[5] = {1, 2, 3, 4, 5};
    int b[5] = {10, 20, 30, 40, 50};
    int c[5];
    
    int *d_a, *d_b, *d_c; // GPU copies of a, b, c
    
    int size = 5 * sizeof(int);
    
    // Allocate space for device copies of a, b, c
    cudaMalloc((void **)&d_a, size);
    cudaMalloc((void **)&d_b, size);
    cudaMalloc((void **)&d_c, size);
    
    // Copy inputs to the device
    cudaMemcpy(d_a, a, size, cudaMemcpyHostToDevice);
    cudaMemcpy(d_b, b, size, cudaMemcpyHostToDevice);
    
    // Launch add() kernel on GPU with 5 threads
    add<<<1, 5>>>(d_a, d_b, d_c);
    
    // Copy result back to host
    cudaMemcpy(c, d_c, size, cudaMemcpyDeviceToHost);
    
    // Print the result
    printf("Result: ");
    for (int i = 0; i < 5; i++) {
        printf("%d ", c[i]);
    }
    
    // Cleanup
    cudaFree(d_a);
    cudaFree(d_b);
    cudaFree(d_c);
    
    return 0;
}
\end{lstlisting}

\textbf{Explanation:}

\begin{itemize}
\item We declare and initialize arrays \texttt{a}, \texttt{b}, and \texttt{c}.
\item \texttt{cudaMalloc()} allocates memory on the GPU.
\item \texttt{cudaMemcpy()} is used to copy data from the CPU to the GPU and vice versa.
\item We launch the kernel using the syntax \texttt{add<<<1, 5>>>}, where \texttt{1} indicates one block, and \texttt{5} indicates 5 threads in the block.
\item Finally, we clean up the memory using \texttt{cudaFree()}.
\end{itemize}

\subsection{CUDA Program Code structure}

Writing a basic CUDA program involves several key steps. In this section, we will go over these steps to help you understand how to transfer data between the CPU (host) and the GPU (device), execute a kernel on the GPU, and clean up resources after execution.

Here are the typical steps involved:

\begin{enumerate}
    \item \textbf{Allocate memory on the host and device}: 
    First, you need to allocate memory for both the CPU (host) and GPU (device). The CPU will store the input data, and the GPU will process it.
\begin{lstlisting}[style=cpp, caption={C/C++ Code With CUDA}]
// Host allocation
int *h_data = (int*)malloc(size);

// Device allocation
int *d_data;
cudaMalloc(&d_data, size);
\end{lstlisting}
    
    \item \textbf{Copy data from host to device}:
    After memory is allocated, the next step is to transfer data from the host (CPU) to the device (GPU).
\begin{lstlisting}[style=cpp, caption={C/C++ Code With CUDA}]
cudaMemcpy(d_data, h_data, size, cudaMemcpyHostToDevice);
\end{lstlisting}
    
    \item \textbf{Launch the kernel}: 
    The kernel function, which defines the operations to be executed on the GPU, is then launched. You need to specify the number of threads and blocks.
\begin{lstlisting}[style=cpp, caption={C/C++ Code With CUDA}]
kernel<<<num_blocks, num_threads>>>(d_data);
\end{lstlisting}
    
    \item \textbf{Copy results from device to host}: 
    After the kernel finishes executing on the GPU, you need to copy the results back to the CPU.
\begin{lstlisting}[style=cpp, caption={C/C++ Code With CUDA}]
cudaMemcpy(h_data, d_data, size, cudaMemcpyDeviceToHost);
\end{lstlisting}
    
    \item \textbf{Free the memory}: 
    Finally, it is important to free the memory allocated on both the host and device to prevent memory leaks.
\begin{lstlisting}[style=cpp, caption={C/C++ Code With CUDA}]
free(h_data);
cudaFree(d_data);
\end{lstlisting}
\end{enumerate}

This is a basic structure for writing a simple CUDA program. In more complex applications, you can extend these steps to handle more sophisticated memory management and kernel functions.

\subsection{Compiling and Running CUDA Code}
CUDA programs are compiled using NVIDIA's CUDA compiler, \texttt{nvcc}. Before we proceed, make sure you have the CUDA toolkit installed on your machine. 

To compile a CUDA program, use the following command in your terminal:

\begin{lstlisting}[style=cmd]
nvcc -o add_arrays add_arrays.cu
\end{lstlisting}

This command compiles the CUDA source file \texttt{add\_arrays.cu} and generates an executable named \texttt{add\_arrays}. You can run the executable by typing:

\begin{lstlisting}[style=cmd]
./add_arrays
\end{lstlisting}

If everything is set up correctly, you should see the output:

\begin{lstlisting}[style=cmd]
Result: 11 22 33 44 55
\end{lstlisting}

\subsubsection{Configuring the Environment for CUDA Development}
To compile and run CUDA code, you need to have the CUDA toolkit and NVIDIA drivers installed. Follow the steps below to configure your environment:

\begin{enumerate}
\item \textbf{Install CUDA Toolkit}: Visit the official NVIDIA website and download the latest CUDA toolkit for your operating system. Follow the instructions provided by NVIDIA for installation.
\item \textbf{Check GPU Compatibility}: Ensure that your system has a CUDA-enabled GPU. You can check this by running:
\begin{lstlisting}[style=cmd]
nvidia-smi
\end{lstlisting}
This command displays details about your GPU, including whether it supports CUDA.
\item \textbf{Set Environment Variables}: After installation, you may need to add the CUDA binaries to your system's PATH. For example, on Linux, add the following lines to your \texttt{.bashrc} file:
\begin{lstlisting}[style=cmd]
export PATH=/usr/local/cuda/bin:$PATH
export LD_LIBRARY_PATH=/usr/local/cuda/lib64:$LD_LIBRARY_PATH
\end{lstlisting}
On Windows, you can add these paths through the System Properties window.
\end{enumerate}

\subsection{Conclusion}
Writing your first CUDA program involves understanding the basics of parallelism, memory management, and thread execution. Once your environment is set up, you can begin experimenting with more complex problems and optimizing your code for performance on the GPU. By distributing tasks over thousands of threads, CUDA enables massive parallelism and accelerates many types of computation.

\section{Basic GPU Algorithms}
In this section, we will explore some simple algorithms that can be efficiently implemented on a GPU. These algorithms introduce fundamental concepts like parallelism and thread management. To help you understand these algorithms better, we'll walk through detailed examples, using CUDA as the primary framework.

\subsection{Vector Addition: The Fundamentals}
One of the simplest operations you can perform on a GPU is vector addition. This involves adding two arrays element-wise. The advantage of using a GPU is that you can perform multiple operations in parallel, which can drastically improve performance.

In CUDA, we use threads to perform the addition of the vectors. Each thread handles the computation for one element in the vector. The idea is to map each thread to one data element, allowing the GPU to compute the sum of the entire array in parallel.

Let’s assume we have two vectors \(A\) and \(B\), each with \(N\) elements. The task is to add them together to produce a third vector \(C\), where \(C[i] = A[i] + B[i]\).

The following is a simple CUDA kernel for vector addition:

\begin{lstlisting}[style=cpp, caption={C/C++ Code With CUDA}]
// CUDA Kernel for vector addition
__global__ void vectorAdd(float *A, float *B, float *C, int N) {
    int i = threadIdx.x + blockIdx.x * blockDim.x;
    if (i < N) {
        C[i] = A[i] + B[i];
    }
}
\end{lstlisting}

\textbf{Explanation}:

\begin{itemize}
\item \textbf{Threads and Blocks}: The computation is parallelized by assigning each thread to compute one element. `threadIdx.x` represents the thread's index within a block, and `blockIdx.x` is the index of the block itself. `blockDim.x` tells us how many threads there are in each block. Together, these give the global index \(i\), which corresponds to an element in the array.
\item \textbf{Boundary Checking}: We check `if (i < N)` to ensure that threads do not access elements beyond the array’s bounds.
\end{itemize}

The host (CPU) side code to invoke this kernel is as follows:

\begin{lstlisting}[style=cpp, caption={C/C++ Code With CUDA}]
// Host code to allocate memory and call the kernel
int N = 1024;
size_t size = N * sizeof(float);

float *h_A = (float *)malloc(size);
float *h_B = (float *)malloc(size);
float *h_C = (float *)malloc(size);

// Initialize vectors A and B
for (int i = 0; i < N; i++) {
    h_A[i] = i;
    h_B[i] = i * 2;
}

float *d_A, *d_B, *d_C;
cudaMalloc(&d_A, size);
cudaMalloc(&d_B, size);
cudaMalloc(&d_C, size);

// Copy vectors from host memory to device memory
cudaMemcpy(d_A, h_A, size, cudaMemcpyHostToDevice);
cudaMemcpy(d_B, h_B, size, cudaMemcpyHostToDevice);

// Launch the kernel with N/256 blocks and 256 threads per block
int threadsPerBlock = 256;
int blocksPerGrid = (N + threadsPerBlock - 1) / threadsPerBlock;
vectorAdd<<<blocksPerGrid, threadsPerBlock>>>(d_A, d_B, d_C, N);

// Copy the result from device to host
cudaMemcpy(h_C, d_C, size, cudaMemcpyDeviceToHost);

// Free device memory
cudaFree(d_A);
cudaFree(d_B);
cudaFree(d_C);
\end{lstlisting}

In this example, we initialize two vectors on the host, copy them to the GPU, execute the vector addition in parallel, and then copy the result back to the host.

\subsection{Summing Arrays: Parallel Reduction}

Summing a large array is another common problem that benefits from parallelization on a GPU. The basic idea behind parallel reduction is to split the array into smaller chunks and sum them in parallel.

For example, given an array of \(N\) elements, you can have \(N/2\) threads each sum two elements, reducing the problem size by half. This process repeats until there’s only one element left—the sum of the entire array.

\paragraph{Summing Arrays: Parallel Reduction Example}

Consider an array of 16 random numbers. The array is reduced by summing pairs of elements in parallel until a single sum remains. The process is as follows:

\[
\text{Array} = [8, 3, 5, 7, 2, 9, 1, 6, 4, 10, 12, 15, 11, 14, 13, 16]
\]

The steps are shown below, where each row represents one step of the reduction process.

\[
\begin{array}{|c|c|}
\hline
\text{Step} & \text{Array} \\
\hline
\text{Initial Array} & [8, 3, 5, 7, 2, 9, 1, 6, 4, 10, 12, 15, 11, 14, 13, 16] \\
\hline
1 & [8+3, 5+7, 2+9, 1+6, 4+10, 12+15, 11+14, 13+16] \\
  & [11, 12, 11, 7, 14, 27, 25, 29] \\
\hline
2 & [11+12, 11+7, 14+27, 25+29] \\
  & [23, 18, 41, 54] \\
\hline
3 & [23+18, 41+54] \\
  & [41, 95] \\
\hline
4 & [41+95] \\
  & [136] \\
\hline
\end{array}
\]

In this example, the array is reduced from 16 elements down to 8, then to 4, then to 2, and finally to 1, which is the sum of the entire array. At each step, pairs of elements are summed in parallel.

The final sum of the array is 136.

\textcolor{codegreen}{\textbf{In actual GPU threads,}} each column corresponds to a thread, and the table shows how the values change at each step of the reduction. Empty cells indicate that the thread has completed its task in that step and is no longer active.

\[
\begin{tabularx}{\textwidth}{|c|X|X|X|X|X|X|X|X|X|X|X|X|X|X|X|X|}
\hline
\text{Step} & $T_1$ & $T_2$ & $T_3$ & $T_4$ & $T_5$ & $T_6$ & $T_7$ & $T_8$ & $T_9$ & $T_{10}$ & $T_{11}$ & $T_{12}$ & $T_{13}$ & $T_{14}$ & $T_{15}$ & $T_{16}$ \\
\hline
\text{Initial} & 8 & 3 & 5 & 7 & 2 & 9 & 1 & 6 & 4 & 10 & 12 & 15 & 11 & 14 & 13 & 16 \\
\hline
\text{Step 1} & 11 &  & 12 &  & 11 &  & 7 &  & 14 &  & 27 &  & 25 &  & 29 &  \\
\hline
\text{Step 2} & 23 &  &  &  & 18 &  &  &  & 41 &  &  &  & 54 &  &  & \\
\hline
\text{Step 3} & 41 &  &  &  &  &  &  &  & 95 &  &  &  &  &  &  & \\
\hline
\text{Step 4} & 136 &  &  &  &  &  &  &  &  &  &  &  &  &  &  & \\
\hline
\end{tabularx}
\]

In this representation:
\begin{itemize}
    \item Each thread \( T_i \) corresponds to a column, and values in the array are reduced through summing adjacent pairs.
    \item In Step 1, each thread sums two adjacent elements (e.g., \( T_1 \) sums 8 and 3 to get 11, and so on). Empty columns represent threads that are idle in that step.
    \item In Step 2, the active threads sum the results from Step 1 in pairs.
    \item This process continues until only one thread remains with the final sum.
\end{itemize}

\paragraph{Here is a simple kernel for parallel reduction:}

CUDA Code is shown below.

\begin{lstlisting}[style=cpp, caption={C/C++ Code With CUDA}]
// CUDA Kernel for parallel reduction
__global__ void reduceSum(float *input, float *output, int N) {
    extern __shared__ float sdata[];
    int tid = threadIdx.x;
    int i = blockIdx.x * blockDim.x + tid;
    
    // Load input into shared memory
    sdata[tid] = (i < N) ? input[i] : 0.0f;
    __syncthreads();

    // Perform reduction in shared memory
    for (int s = blockDim.x / 2; s > 0; s >>= 1) {
        if (tid < s) {
            sdata[tid] += sdata[tid + s];
        }
        __syncthreads();
    }

    // Write result for this block to global memory
    if (tid == 0) {
        output[blockIdx.x] = sdata[0];
    }
}
\end{lstlisting}

\textbf{Explanation}:

\begin{itemize}
\item \textbf{Shared Memory}: We use shared memory to store the intermediate results of the reduction within each block. Shared memory is much faster than global memory.
\item \textbf{Synchronization}: The `\_\_syncthreads()' function ensures that all threads have finished their work before proceeding to the next step of the reduction.
\item \textbf{Reduction Loop}: In each iteration of the loop, the number of elements to be summed is halved, and threads with indices less than half the block size continue summing.
\end{itemize}

The process continues until each block has produced a partial sum, which can be further reduced by another kernel call or using the CPU.

\subsection{Cumulative Sum (Prefix Sum)}

The prefix sum (or cumulative sum) is a widely used parallel algorithm, especially for applications like sorting and graph traversal. The goal of the prefix sum is to produce an output array where each element is the sum of all previous elements in the input array.

\textbf{Inclusive Scan}: In an inclusive scan, the element at position \(i\) in the output array is the sum of all elements up to and including position \(i\) in the input array. 

\textbf{Exclusive Scan}: In an exclusive scan, the element at position \(i\) in the output array is the sum of all elements before position \(i\) in the input array.

\paragraph{Cumulative Sum (Prefix Sum) Example}

Given the array:

\[
\text{Array} = [8,\, 3,\, 5,\, 7,\, 2,\, 9,\, 1,\, 6,\, 4,\, 10,\, 12,\, 15,\, 11,\, 14,\, 13,\, 16]
\]

We will perform an \textbf{inclusive scan}, where each element in the output array is the sum of all elements up to and including that position in the input array.

\subsection*{Step-by-Step Calculation}

\begin{enumerate}
    \item \textbf{Initial Array}:
    \[
    [8,\, 3,\, 5,\, 7,\, 2,\, 9,\, 1,\, 6,\, 4,\, 10,\, 12,\, 15,\, 11,\, 14,\, 13,\, 16]
    \]
    
    \item \textbf{Cumulative Sum Calculation}:
    \begin{align*}
    \text{Position }0 &: \quad 8 \\
    \text{Position }1 &: \quad 8 + 3 = 11 \\
    \text{Position }2 &: \quad 11 + 5 = 16 \\
    \text{Position }3 &: \quad 16 + 7 = 23 \\
    \text{Position }4 &: \quad 23 + 2 = 25 \\
    \text{Position }5 &: \quad 25 + 9 = 34 \\
    \text{Position }6 &: \quad 34 + 1 = 35 \\
    \text{Position }7 &: \quad 35 + 6 = 41 \\
    \text{Position }8 &: \quad 41 + 4 = 45 \\
    \text{Position }9 &: \quad 45 + 10 = 55 \\
    \text{Position }10 &: \quad 55 + 12 = 67 \\
    \text{Position }11 &: \quad 67 + 15 = 82 \\
    \text{Position }12 &: \quad 82 + 11 = 93 \\
    \text{Position }13 &: \quad 93 + 14 = 107 \\
    \text{Position }14 &: \quad 107 + 13 = 120 \\
    \text{Position }15 &: \quad 120 + 16 = 136 \\
    \end{align*}
    
    \item \textbf{Final Cumulative Sum Array}:
    \[
    \text{Cumulative Sum Array} = [8,\, 11,\, 16,\, 23,\, 25,\, 34,\, 35,\, 41,\, 45,\, 55,\, 67,\, 82,\, 93,\, 107,\, 120,\, 136]
    \]
\end{enumerate}

Now, we can represent the process in terms of GPU threads, where each column corresponds to a thread. The table below shows how the values are updated at each step of the cumulative sum process.

\paragraph{Parallel Calculation Using GPU Threads}

To illustrate how this process can be parallelized using GPU threads, we can use a standard parallel prefix sum algorithm. One common method is the \textbf{Hillis-Steele scan}, which updates the array in logarithmic steps. Here's how it works for our array of 16 elements.

\subsubsection*{Algorithm Overview}

\begin{itemize}
    \item \textbf{Step \( d = 0 \)} (distance \( = 1 \)):
    \[
    \text{For all positions } i \geq 1, \quad A[i] = A[i] + A[i - 1]
    \]
    \item \textbf{Step \( d = 1 \)} (distance \( = 2 \)):
    \[
    \text{For all positions } i \geq 2, \quad A[i] = A[i] + A[i - 2]
    \]
    \item \textbf{Step \( d = 2 \)} (distance \( = 4 \)):
    \[
    \text{For all positions } i \geq 4, \quad A[i] = A[i] + A[i - 4]
    \]
    \item \textbf{Step \( d = 3 \)} (distance \( = 8 \)):
    \[
    \text{For all positions } i \geq 8, \quad A[i] = A[i] + A[i - 8]
    \]
\end{itemize}

\subsubsection*{Detailed Steps with GPU Threads}

We will represent each element as being processed by a separate thread \( T_i \), where \( i \) ranges from 1 to 16.

\paragraph{Initial State}

\begin{center}
\begin{tabular}{|c|*{16}{c|}}
\hline
\textbf{Thread} & \( T_1 \) & \( T_2 \) & \( T_3 \) & \( T_4 \) & \( T_5 \) & \( T_6 \) & \( T_7 \) & \( T_8 \) & \( T_9 \) & \( T_{10} \) & \( T_{11} \) & \( T_{12} \) & \( T_{13} \) & \( T_{14} \) & \( T_{15} \) & \( T_{16} \) \\
\hline
\textbf{Value} & 8 & 3 & 5 & 7 & 2 & 9 & 1 & 6 & 4 & 10 & 12 & 15 & 11 & 14 & 13 & 16 \\
\hline
\end{tabular}
\end{center}

\paragraph{Step 1 (\( d = 0 \), distance \( = 1 \))}

\textbf{Operation}:
\[
\text{If } i \geq 2, \quad \text{Value}_i = \text{Value}_i + \text{Value}_{i - 1}
\]

\textbf{Updated Values}:

\begin{align*}
T_1 &: \quad 8 \quad (\text{unchanged}) \\
T_2 &: \quad 3 + 8 = 11 \\
T_3 &: \quad 5 + 3 = 8 \\
T_4 &: \quad 7 + 5 = 12 \\
T_5 &: \quad 2 + 7 = 9 \\
T_6 &: \quad 9 + 2 = 11 \\
T_7 &: \quad 1 + 9 = 10 \\
T_8 &: \quad 6 + 1 = 7 \\
T_9 &: \quad 4 + 6 = 10 \\
T_{10} &: \quad 10 + 4 = 14 \\
T_{11} &: \quad 12 + 10 = 22 \\
T_{12} &: \quad 15 + 12 = 27 \\
T_{13} &: \quad 11 + 15 = 26 \\
T_{14} &: \quad 14 + 11 = 25 \\
T_{15} &: \quad 13 + 14 = 27 \\
T_{16} &: \quad 16 + 13 = 29 \\
\end{align*}

\textbf{Resulting Values}:

\begin{center}
\begin{tabular}{|c|*{16}{c|}}
\hline
\textbf{Thread} & 8 & 11 & 8 & 12 & 9 & 11 & 10 & 7 & 10 & 14 & 22 & 27 & 26 & 25 & 27 & 29 \\
\hline
\end{tabular}
\end{center}

\paragraph{Step 2 (\( d = 1 \), distance \( = 2 \))}

\textbf{Operation}:
\[
\text{If } i \geq 3, \quad \text{Value}_i = \text{Value}_i + \text{Value}_{i - 2}
\]

\textbf{Updated Values}:

\begin{align*}
T_1 &: \quad 8 \quad (\text{unchanged}) \\
T_2 &: \quad 11 \quad (\text{unchanged}) \\
T_3 &: \quad 8 + 8 = 16 \\
T_4 &: \quad 12 + 11 = 23 \\
T_5 &: \quad 9 + 8 = 17 \\
T_6 &: \quad 11 + 12 = 23 \\
T_7 &: \quad 10 + 9 = 19 \\
T_8 &: \quad 7 + 11 = 18 \\
T_9 &: \quad 10 + 10 = 20 \\
T_{10} &: \quad 14 + 7 = 21 \\
T_{11} &: \quad 22 + 10 = 32 \\
T_{12} &: \quad 27 + 14 = 41 \\
T_{13} &: \quad 26 + 22 = 48 \\
T_{14} &: \quad 25 + 27 = 52 \\
T_{15} &: \quad 27 + 26 = 53 \\
T_{16} &: \quad 29 + 25 = 54 \\
\end{align*}

\textbf{Resulting Values}:

\begin{center}
\begin{tabular}{|c|*{16}{c|}}
\hline
\textbf{Thread} & 8 & 11 & 16 & 23 & 17 & 23 & 19 & 18 & 20 & 21 & 32 & 41 & 48 & 52 & 53 & 54 \\
\hline
\end{tabular}
\end{center}

\paragraph{Step 3 (\( d = 2 \), distance \( = 4 \))}

\textbf{Operation}:
\[
\text{If } i \geq 5, \quad \text{Value}_i = \text{Value}_i + \text{Value}_{i - 4}
\]

\textbf{Updated Values}:

\begin{align*}
T_1 &: \quad 8 \quad (\text{unchanged}) \\
T_2 &: \quad 11 \quad (\text{unchanged}) \\
T_3 &: \quad 16 \quad (\text{unchanged}) \\
T_4 &: \quad 23 \quad (\text{unchanged}) \\
T_5 &: \quad 17 + 8 = 25 \\
T_6 &: \quad 23 + 11 = 34 \\
T_7 &: \quad 19 + 16 = 35 \\
T_8 &: \quad 18 + 23 = 41 \\
T_9 &: \quad 20 + 17 = 37 \\
T_{10} &: \quad 21 + 23 = 44 \\
T_{11} &: \quad 32 + 19 = 51 \\
T_{12} &: \quad 41 + 18 = 59 \\
T_{13} &: \quad 48 + 20 = 68 \\
T_{14} &: \quad 52 + 21 = 73 \\
T_{15} &: \quad 53 + 32 = 85 \\
T_{16} &: \quad 54 + 41 = 95 \\
\end{align*}

\textbf{Resulting Values}:

\begin{center}
\begin{tabular}{|c|*{16}{c|}}
\hline
\textbf{Thread} & 8 & 11 & 16 & 23 & 25 & 34 & 35 & 41 & 37 & 44 & 51 & 59 & 68 & 73 & 85 & 95 \\
\hline
\end{tabular}
\end{center}

\paragraph{Step 4 (\( d = 3 \), distance \( = 8 \))}

\textbf{Operation}:
\[
\text{If } i \geq 9, \quad \text{Value}_i = \text{Value}_i + \text{Value}_{i - 8}
\]

\textbf{Updated Values}:

\begin{align*}
T_1 &: \quad 8 \quad (\text{unchanged}) \\
T_2 &: \quad 11 \quad (\text{unchanged}) \\
T_3 &: \quad 16 \quad (\text{unchanged}) \\
T_4 &: \quad 23 \quad (\text{unchanged}) \\
T_5 &: \quad 25 \quad (\text{unchanged}) \\
T_6 &: \quad 34 \quad (\text{unchanged}) \\
T_7 &: \quad 35 \quad (\text{unchanged}) \\
T_8 &: \quad 41 \quad (\text{unchanged}) \\
T_9 &: \quad 37 + 8 = 45 \\
T_{10} &: \quad 44 + 11 = 55 \\
T_{11} &: \quad 51 + 16 = 67 \\
T_{12} &: \quad 59 + 23 = 82 \\
T_{13} &: \quad 68 + 25 = 93 \\
T_{14} &: \quad 73 + 34 = 107 \\
T_{15} &: \quad 85 + 35 = 120 \\
T_{16} &: \quad 95 + 41 = 136 \\
\end{align*}

\textbf{Resulting Values (Final Cumulative Sum)}:

\begin{center}
\begin{tabular}{|c|*{16}{c|}}
\hline
\textbf{Thread} & 8 & 11 & 16 & 23 & 25 & 34 & 35 & 41 & 45 & 55 & 67 & 82 & 93 & 107 & 120 & 136 \\
\hline
\end{tabular}
\end{center}

\paragraph{Summary}

The cumulative sum (prefix sum) process results in the array:

\[
[8,\, 11,\, 16,\, 23,\, 25,\, 34,\, 35,\, 41,\, 45,\, 55,\, 67,\, 82,\, 93,\, 107,\, 120,\, 136]
\]

\paragraph{Here’s a basic kernel for an exclusive scan using the work-efficient scan algorithm:}

CUDA Code is shown below.

\begin{lstlisting}[style=cpp, caption={C/C++ Code With CUDA}]
// CUDA Kernel for exclusive scan (prefix sum)
__global__ void exclusiveScan(float *input, float *output, int N) {
    extern __shared__ float temp[]; // Shared memory
    int tid = threadIdx.x;
    
    // Load input into shared memory
    temp[2 * tid] = input[2 * tid];
    temp[2 * tid + 1] = input[2 * tid + 1];
    __syncthreads();
    
    // Up-sweep (reduce) phase
    for (int stride = 1; stride <= blockDim.x; stride *= 2) {
        int index = (tid + 1) * stride * 2 - 1;
        if (index < N) {
            temp[index] += temp[index - stride];
        }
        __syncthreads();
    }
    
    // Down-sweep phase
    for (int stride = blockDim.x / 2; stride > 0; stride /= 2) {
        int index = (tid + 1) * stride * 2 - 1;
        if (index < N) {
            float temp_val = temp[index - stride];
            temp[index - stride] = temp[index];
            temp[index] += temp_val;
        }
        __syncthreads();
    }
    
    // Write results to output
    output[2 * tid] = temp[2 * tid];
    output[2 * tid + 1] = temp[2 * tid + 1];
}
\end{lstlisting}

\textbf{Explanation}:

\begin{itemize}
\item \textbf{Up-Sweep (Reduction)}: The algorithm first performs a reduction where each element is added to its corresponding stride element. This builds up the sums.
\item \textbf{Down-Sweep (Distribution)}: After the reduction, a down-sweep distributes the sums to the appropriate elements, resulting in the final prefix sum.
\item \textbf{Shared Memory}: The algorithm uses shared memory for fast access to intermediate results.
\end{itemize}

This concludes our introduction to some basic GPU algorithms. By understanding these simple yet powerful algorithms, you can leverage the parallel nature of GPUs to greatly speed up various computations.

\section{Matrix Operations}

Matrix operations are fundamental in various scientific applications, especially in fields like machine learning, computer vision, and scientific computing. In this section, we will explore key matrix operations and how they can be parallelized to improve efficiency using Python. Specifically, we will discuss matrix addition and matrix multiplication, showing how to leverage Python's multi-threading capabilities, CUDA programming for GPU acceleration, and optimized libraries to speed up these operations.

\subsection{Matrix Addition}

Matrix addition is one of the simplest matrix operations. It involves adding corresponding elements of two matrices to produce a new matrix. Given two matrices \( A \) and \( B \), their sum \( C \) is calculated as:

\[
C_{ij} = A_{ij} + B_{ij}
\]

where \( i \) and \( j \) are the row and column indices of the matrices.

\subsubsection{Parallel Implementation of Matrix Addition}

Matrix addition is an element-wise operation, which makes it highly parallelizable. Each element in the resulting matrix can be computed independently. Using Python, we can parallelize this operation using the \texttt{concurrent.futures} module, \texttt{multiprocessing}, or even GPU acceleration using CUDA.

\paragraph{Example Code for Sequential Matrix Addition}

Before diving into parallelism, let's look at how matrix addition is implemented sequentially:

\begin{lstlisting}[style=python]
import numpy as np

# Define two matrices
A = np.array([[1, 2, 3], 
              [4, 5, 6], 
              [7, 8, 9]])

B = np.array([[9, 8, 7], 
              [6, 5, 4], 
              [3, 2, 1]])

# Sequential matrix addition
C = A + B

print(C)
\end{lstlisting}

\paragraph{Parallel Matrix Addition Using Threads}

For parallelizing the matrix addition, we can divide the matrix into rows or blocks and assign each portion to a separate thread for computation. Here's how you can do it using the \texttt{ThreadPoolExecutor} from the \texttt{concurrent.futures} module:

\begin{lstlisting}[style=python]
import numpy as np
from concurrent.futures import ThreadPoolExecutor

# Function to add two matrices row by row
def add_rows(row_a, row_b):
    return row_a + row_b

# Matrix addition with threading
def parallel_matrix_add(A, B):
    with ThreadPoolExecutor() as executor:
        result = list(executor.map(add_rows, A, B))
    return np.array(result)

A = np.array([[1, 2, 3], 
              [4, 5, 6], 
              [7, 8, 9]])

B = np.array([[9, 8, 7], 
              [6, 5, 4], 
              [3, 2, 1]])

# Perform parallel matrix addition
C = parallel_matrix_add(A, B)
print(C)
\end{lstlisting}

\paragraph{Matrix Addition Using CUDA}

We can further optimize matrix addition by leveraging CUDA for GPU acceleration. CUDA enables the use of GPUs to perform parallel matrix operations on a massive scale, making it much faster for large matrices. Below is an example of using CUDA for matrix addition:

\begin{lstlisting}[style=python]
import numpy as np
from numba import cuda

# Define the matrix size
N = 3

# CUDA kernel for matrix addition
@cuda.jit
def matrix_add_cuda(A, B, C):
    i, j = cuda.grid(2)
    if i < C.shape[0] and j < C.shape[1]:
        C[i, j] = A[i, j] + B[i, j]

# Initialize matrices
A = np.array([[1, 2, 3], 
              [4, 5, 6], 
              [7, 8, 9]]).astype(np.float32)

B = np.array([[9, 8, 7], 
              [6, 5, 4], 
              [3, 2, 1]]).astype(np.float32)

C = np.zeros((N, N), dtype=np.float32)

# Define grid and block size
threads_per_block = (16, 16)
blocks_per_grid_x = int(np.ceil(A.shape[0] / threads_per_block[0]))
blocks_per_grid_y = int(np.ceil(A.shape[1] / threads_per_block[1]))
blocks_per_grid = (blocks_per_grid_x, blocks_per_grid_y)

# Call the CUDA kernel
matrix_add_cuda[blocks_per_grid, threads_per_block](A, B, C)

print(C)
\end{lstlisting}

\paragraph{Explanation:}

In the CUDA implementation, the \texttt{matrix\_add\_cuda} function runs in parallel on the GPU. The grid and block dimensions control how the matrix is split across the GPU threads.

\subsection{Matrix Multiplication: Naive, Optimized, and CUDA Approaches}

Matrix multiplication is a more complex operation than matrix addition. Given two matrices \( A \) of dimensions \( m \times n \) and \( B \) of dimensions \( n \times p \), their product \( C \) is computed as:

\[
C_{ij} = \sum_{k=1}^{n} A_{ik} \times B_{kj}
\]

\subsubsection{Naive Implementation of Matrix Multiplication}

Here is the naive Python implementation of matrix multiplication:

\begin{lstlisting}[style=python]
import numpy as np

A = np.array([[1, 2, 3], 
              [4, 5, 6]])

B = np.array([[7, 8], 
              [9, 10], 
              [11, 12]])

# Naive matrix multiplication
C = np.zeros((A.shape[0], B.shape[1]))

for i in range(A.shape[0]):
    for j in range(B.shape[1]):
        for k in range(A.shape[1]):
            C[i, j] += A[i, k] * B[k, j]

print(C)
\end{lstlisting}

\subsubsection{Naive Matrix Multiplication}

The naive approach to matrix multiplication involves three nested loops: one for rows of matrix \( A \), one for columns of matrix \( B \), and one for summing the element-wise products. Here’s how you can implement this in Python:

\begin{lstlisting}[style=python]
import numpy as np

# Naive matrix multiplication
def naive_matrix_multiply(A, B):
    m, n = A.shape
    n, p = B.shape
    C = np.zeros((m, p))

    for i in range(m):
        for j in range(p):
            for k in range(n):
                C[i, j] += A[i, k] * B[k, j]

    return C

# Define two matrices
A = np.array([[1, 2], 
              [3, 4]])

B = np.array([[5, 6], 
              [7, 8]])

C = naive_matrix_multiply(A, B)
print(C)
\end{lstlisting}

\subsubsection{Optimized Matrix Multiplication}

The naive matrix multiplication is inefficient because it repeatedly reads the same data from memory, causing memory latency issues. A more optimized approach involves using \textbf{shared memory} and leveraging matrix libraries like \textbf{NumPy}, which are highly optimized and make use of \textbf{BLAS (Basic Linear Algebra Subprograms)} \cite{blackford2002updated}.

\paragraph{Using NumPy for Optimized Multiplication}

NumPy’s \texttt{dot} function is a highly optimized implementation of matrix multiplication:

\begin{lstlisting}[style=python]
import numpy as np

# Define two matrices
A = np.array([[1, 2], 
              [3, 4]])

B = np.array([[5, 6], 
              [7, 8]])

# Using NumPy's optimized dot product for matrix multiplication
C = np.dot(A, B)
print(C)
\end{lstlisting}

NumPy uses highly optimized libraries under the hood (like OpenBLAS or Intel MKL) to perform matrix multiplication efficiently, taking advantage of low-level optimizations such as data pre-fetching and cache reuse.

\subsubsection{Parallelizing Matrix Multiplication}

To parallelize matrix multiplication manually, we can break the operation down by assigning different rows of matrix \( A \) and different columns of matrix \( B \) to different threads. However, using optimized libraries like NumPy is often the best approach, as they already include multi-threading and SIMD (Single Instruction, Multiple Data) optimizations.

Here’s an example of manually parallelizing matrix multiplication:

\begin{lstlisting}[style=python]
from concurrent.futures import ThreadPoolExecutor
import numpy as np

# Function to compute one row of matrix C
def multiply_row(A_row, B):
    return np.dot(A_row, B)

# Parallel matrix multiplication
def parallel_matrix_multiply(A, B):
    m = A.shape[0]
    C = np.zeros((m, B.shape[1]))

    with ThreadPoolExecutor() as executor:
        results = list(executor.map(multiply_row, A, [B] * m))

    return np.array(results)

# Define matrices
A = np.array([[1, 2], 
              [3, 4]])

B = np.array([[5, 6], 
              [7, 8]])

# Perform parallel matrix multiplication
C = parallel_matrix_multiply(A, B)
print(C)
\end{lstlisting}

\paragraph{Explanation:}

In this implementation, each thread computes a row of the result matrix \( C \) by performing the dot product of a row from \( A \) with matrix \( B \).

\subsubsection{Matrix Multiplication Using CUDA}

Matrix multiplication can be significantly accelerated using CUDA. Here's how you can implement matrix multiplication on the GPU using CUDA:

\begin{lstlisting}[style=python]
import numpy as np
from numba import cuda

# CUDA kernel for matrix multiplication
@cuda.jit
def matrix_multiply_cuda(A, B, C):
    row, col = cuda.grid(2)
    if row < C.shape[0] and col < C.shape[1]:
        tmp = 0
        for k in range(A.shape[1]):
            tmp += A[row, k] * B[k, col]
        C[row, col] = tmp

# Initialize matrices
A = np.array([[1, 2, 3], 
              [4, 5, 6]]).astype(np.float32)

B = np.array([[7, 8], 
              [9, 10], 
              [11, 12]]).astype(np.float32)

C = np.zeros((A.shape[0], B.shape[1]), dtype=np.float32)

# Define grid and block size
threads_per_block = (16, 16)
blocks_per_grid_x = int(np.ceil(A.shape[0] / threads_per_block[0]))
blocks_per_grid_y = int(np.ceil(B.shape[1] / threads_per_block[1]))
blocks_per_grid = (blocks_per_grid_x, blocks_per_grid_y)

# Call the CUDA kernel
matrix_multiply_cuda[blocks_per_grid, threads_per_block](A, B, C)

print(C)
\end{lstlisting}

\paragraph{Explanation:}

In the CUDA version of matrix multiplication, the matrices are divided into blocks and threads, allowing parallel execution of multiplication across rows and columns on the GPU.

\subsubsection{Comparison of Naive, Optimized, and CUDA Approaches}

\begin{formatteditem}
\item\textbf{Naive Approach:} Easy to understand and implement but not efficient for large matrices due to memory latency and redundant memory access. It operates in \(O(n^3)\) time complexity and does not utilize the full potential of the hardware.
  
\item \textbf{Optimized Approach:} Libraries like NumPy leverage low-level optimizations (SIMD instructions, multi-threading) and shared memory to minimize memory latency, making them highly efficient for large-scale matrix multiplication. These approaches improve computational efficiency, but are still limited by the CPU's processing power and memory bandwidth.
    
\item \textbf{CUDA Approach:} Using CUDA, matrix operations can be massively parallelized by distributing the work across thousands of GPU threads. CUDA optimizes memory access patterns and minimizes latency by using shared memory and registers, leading to significant performance gains, especially for large matrices. This approach is highly efficient in handling large datasets and is particularly useful in GPU-intensive applications like machine learning and scientific computing.
\end{formatteditem}

By using optimized libraries, manually parallelizing matrix multiplication, or leveraging GPU acceleration with CUDA, we can significantly improve the performance of matrix operations in Python. These optimizations are crucial for applications involving large datasets, where computation time and efficiency are critical.

\subsubsection{Advanced Matrix Multiplication Algorithms: Strassen and Beyond}

While the parallel matrix multiplication approach we discussed earlier operates in \(O(n^3)\) time complexity, more advanced algorithms have been developed to improve the efficiency of matrix multiplication, particularly for large matrices.

\paragraph{Strassen's Algorithm:}
Strassen's algorithm, introduced by Volker Strassen in 1969, was the first matrix multiplication algorithm that broke the \(O(n^3)\) time complexity barrier. It reduces the number of scalar multiplications required by using a divide-and-conquer approach, achieving a time complexity of approximately \(O(n^{2.81})\). While it offers significant improvements for large matrices, Strassen's algorithm introduces complexity in terms of memory usage and is less efficient for smaller matrices due to its overhead.

\paragraph{Coppersmith-Winograd Algorithm:}
An even faster theoretical approach is the Coppersmith-Winograd algorithm, which achieves a time complexity of \(O(n^{2.376})\). However, this algorithm is not commonly used in practice due to its complexity and large constant factors, making it impractical for real-world applications despite its improved asymptotic performance.

\paragraph{The Current Fastest Method:}
In 2020, researchers developed a new algorithm that further reduced the time complexity to approximately \(O(n^{2.3728596})\). While this is the fastest known theoretical method for matrix multiplication, like the Coppersmith-Winograd algorithm, it remains highly impractical for most real-world applications due to its intricate operations and large memory overhead.

\paragraph{Most Commonly Used Method in Practice:}
Despite the existence of faster theoretical methods, the most commonly used algorithms for matrix multiplication in practice are still variants of the classical \(O(n^3)\) method and Strassen’s algorithm. The classical method is particularly efficient when used in highly optimized libraries like BLAS (Basic Linear Algebra Subprograms), which are widely adopted in computational software.

These libraries optimize the classical algorithm using techniques such as:
- \textbf{Blocking:} Dividing matrices into smaller blocks to better utilize CPU cache and reduce memory access latency.
- \textbf{SIMD (Single Instruction, Multiple Data) instructions:} Leveraging hardware-level parallelism to perform multiple arithmetic operations simultaneously.
- \textbf{Multithreading and GPU acceleration:} Distributing matrix operations across multiple CPU cores or GPUs (e.g., using CUDA).

For smaller matrices or situations where simplicity and portability are key, the classical method optimized with these techniques tends to be the most practical and efficient approach. In contrast, Strassen's algorithm and its variants may be used in large-scale scientific computing and machine learning applications when dealing with very large matrices.

In conclusion, while more advanced algorithms like the Coppersmith-Winograd algorithm and the latest \(O(n^{2.3728596})\) method push the boundaries of theoretical performance, the classical algorithm, along with Strassen’s algorithm, remains the most commonly used due to their simplicity and ease of optimization in practical computational environments.

\section{Optimizing Algorithms for GPU}

When writing algorithms that run on GPUs, especially using CUDA, it is crucial to optimize performance by focusing on memory access patterns, thread synchronization, and making the most of the GPU hardware. In this section, we will cover the essential strategies for achieving better performance by addressing these key areas.

\subsection{Memory Coalescing and Alignment}

Memory coalescing is a technique that ensures efficient memory access by aligning threads' memory requests into fewer transactions. When threads in a warp (a group of 32 threads in CUDA) access consecutive memory addresses, the GPU can combine these accesses into fewer, larger memory transactions. This maximizes memory bandwidth utilization and reduces the overall memory latency.

\textbf{How it works:}
In CUDA, global memory is accessed by all threads, but accessing it efficiently is key to performance. Without proper coalescing, each thread might make its own memory request, leading to multiple, inefficient transactions. With memory coalescing, these requests are combined into a single transaction when:

\begin{itemize}
    \item Threads in a warp access consecutive addresses.
    \item The starting address is properly aligned.
\end{itemize}

\textbf{Ensuring proper alignment:}
To achieve memory coalescing, we need to ensure proper alignment of memory. The memory addresses accessed by each thread should be aligned to the size of the data type. For example, for an array of `float` (4 bytes), the address should be aligned on a 4-byte boundary. This alignment allows the GPU to fetch data in a single coalesced transaction.

An example of memory coalescing with a properly aligned float array:

\begin{lstlisting}[style=cpp, caption={C/C++ Code With CUDA}]
__global__ void coalescedMemoryAccess(float *input, float *output) {
    int tid = threadIdx.x + blockIdx.x * blockDim.x;
    output[tid] = input[tid];  // Each thread accesses a consecutive float
}
\end{lstlisting}

In this case, assuming the `input` and `output` arrays are properly aligned, memory access will be coalesced.

\subsection{Shared Memory Optimization}

Shared memory is a small, fast, on-chip memory that can be used to significantly speed up access times in CUDA programs. Shared memory is accessible by all threads within a block, making it useful for data that needs to be accessed multiple times by different threads.

\textbf{Key advantages of shared memory:}
- Much faster than global memory.
- Reduces redundant global memory accesses.
- Allows efficient data sharing between threads within a block.

To use shared memory effectively, we need to:
1. Load frequently accessed data from global memory into shared memory.
2. Minimize bank conflicts (situations where multiple threads attempt to access the same memory bank simultaneously).

\textbf{Example of using shared memory:}

In this example, we will calculate the sum of an array using shared memory. Each thread block will load a portion of the input array into shared memory, perform the sum, and then write the result back to global memory.

\begin{lstlisting}[style=cpp, caption={C/C++ Code With CUDA}]
__global__ void sumWithSharedMemory(float *input, float *output) {
    __shared__ float shared_data[BLOCK_SIZE];
    
    int tid = threadIdx.x + blockIdx.x * blockDim.x;
    int local_tid = threadIdx.x;
    
    // Load data from global memory into shared memory
    shared_data[local_tid] = input[tid];
    __syncthreads();  // Ensure all threads have loaded their data
    
    // Perform reduction to calculate sum (simple example)
    for (int stride = 1; stride < blockDim.x; stride *= 2) {
        if (local_tid % (2 * stride) == 0) {
            shared_data[local_tid] += shared_data[local_tid + stride];
        }
        __syncthreads();
    }
    
    // Write result back to global memory
    if (local_tid == 0) {
        output[blockIdx.x] = shared_data[0];
    }
}
\end{lstlisting}

\textbf{Bank conflicts:}
CUDA shared memory is divided into banks, and if multiple threads try to access data in the same bank at the same time, a bank conflict occurs, leading to serialization and performance loss. To avoid this, ensure that threads access different memory banks, ideally by organizing data such that consecutive threads access consecutive addresses.

\subsection{Reducing Warp Divergence for Performance}
Warp divergence occurs when threads in a warp follow different execution paths due to conditional branching (e.g., `if-else` statements). When divergence occurs, the GPU must execute both paths serially, which reduces overall performance.

\textbf{How to minimize warp divergence:}
1. Avoid branching whenever possible, or minimize its occurrence by structuring code carefully.
2. Use predication (conditional assignments) instead of branching, which allows all threads to execute the same instruction with different outcomes based on conditions.
3. Ensure branch conditions are uniform across threads in a warp, meaning all threads either take the same branch or none of them do.

\textbf{Example of warp divergence:}

Consider the following code, which results in warp divergence:

\begin{lstlisting}[style=cpp, caption={C/C++ Code With CUDA}]
__global__ void warpDivergenceExample(int *input, int *output) {
    int tid = threadIdx.x;
    if (input[tid] > 0) {
        output[tid] = input[tid] * 2;  // Some threads may execute this
    } else {
        output[tid] = input[tid] / 2;  // Others may execute this
    }
}
\end{lstlisting}

If the condition `input[tid] > 0` evaluates differently for different threads in the same warp, the warp will diverge, and both branches will be executed serially.

To avoid this, consider using predication:

\begin{lstlisting}[style=cpp, caption={C/C++ Code With CUDA}]
__global__ void warpDivergenceAvoided(int *input, int *output) {
    int tid = threadIdx.x;
    int value = input[tid];
    output[tid] = (value > 0) ? value * 2 : value / 2;
}
\end{lstlisting}

Here, instead of branching, the code uses the ternary operator, allowing all threads to follow the same execution path without divergence.

\section{GPU Programming Models Beyond CUDA}

CUDA has become the dominant framework for GPGPU computing, especially for NVIDIA hardware. However, cross-platform GPU programming models like OpenCL, Vulkan Compute, and Metal provide alternatives that can be used with non-NVIDIA hardware such as AMD GPUs or integrated GPUs. This section introduces these programming models, comparing them with CUDA in terms of portability, performance, and ease of use.

\subsection{OpenCL: Cross-Platform GPU Programming}

OpenCL (Open Computing Language) \cite{munshi2009opencl} is an open standard that allows developers to write programs that can run on a variety of platforms, including CPUs, GPUs, FPGAs, and other processors from different vendors.

\subsubsection{Overview of OpenCL}

OpenCL provides a platform-independent API for parallel computing, which makes it a versatile tool for GPGPU programming. Unlike CUDA, which is specific to NVIDIA GPUs, OpenCL works across different hardware vendors like AMD, Intel, and even mobile platforms. OpenCL separates the host (CPU) from the device (GPU), and parallelism is achieved through the concept of kernels.

\subsubsection{OpenCL vs. CUDA}

The primary difference between CUDA and OpenCL lies in their portability. While CUDA is more mature and optimized for NVIDIA hardware, OpenCL offers cross-vendor support, making it suitable for heterogeneous computing environments. However, OpenCL often requires more code and may deliver slightly lower performance compared to CUDA due to its more general-purpose nature.

\begin{formatteditem}
    \item \textbf{Portability}: OpenCL is cross-platform, while CUDA is limited to NVIDIA GPUs.
    \item \textbf{Performance}: CUDA tends to perform better on NVIDIA hardware due to hardware-specific optimizations.
    \item \textbf{Ease of Use}: CUDA offers more user-friendly APIs, while OpenCL requires more manual management of memory and devices.
\end{formatteditem}

\subsubsection{OpenCL Code Example: Vector Addition}

Below is an example of vector addition using OpenCL, showing how a kernel can be defined and executed across multiple platforms:

\begin{verbatim}
__kernel void vector_add(__global const float *A, __global const float *B, __global float *C) {
    int idx = get_global_id(0);
    C[idx] = A[idx] + B[idx];
}

int main() {
    // Platform and device initialization
    cl_platform_id platform;
    cl_device_id device;
    clGetPlatformIDs(1, &platform, NULL);
    clGetDeviceIDs(platform, CL_DEVICE_TYPE_GPU, 1, &device, NULL);
    
    // Memory allocation and buffer creation
    cl_mem d_A = clCreateBuffer(context, CL_MEM_READ_ONLY, size, NULL, NULL);
    cl_mem d_B = clCreateBuffer(context, CL_MEM_READ_ONLY, size, NULL, NULL);
    cl_mem d_C = clCreateBuffer(context, CL_MEM_WRITE_ONLY, size, NULL, NULL);

    // Kernel execution
    clSetKernelArg(kernel, 0, sizeof(cl_mem), (void *)&d_A);
    clSetKernelArg(kernel, 1, sizeof(cl_mem), (void *)&d_B);
    clSetKernelArg(kernel, 2, sizeof(cl_mem), (void *)&d_C);
    clEnqueueNDRangeKernel(queue, kernel, 1, NULL, &global_work_size, NULL, 0, NULL, NULL);
}
\end{verbatim}

\subsection{Vulkan Compute: Low-Level Control for Graphics and Compute}

Vulkan Compute \cite{sellers2016vulkan} is a part of the Vulkan API, designed to provide high-performance, low-level access to graphics and compute hardware. It is particularly suited for graphics-heavy tasks but also supports GPGPU workloads.

\subsubsection{Overview of Vulkan Compute}

Vulkan offers more granular control over the GPU compared to OpenCL and CUDA, which can result in better performance in certain applications. However, this comes at the cost of increased complexity. Vulkan is particularly strong in applications that involve both graphics and compute tasks, as it allows developers to seamlessly switch between rendering and computation.

\subsubsection{Vulkan Compute vs. CUDA}

\begin{formatteditem}
    \item \textbf{Portability}: Vulkan is supported across various platforms (including AMD and Intel GPUs) but requires more complex setup than CUDA.
    \item \textbf{Performance}: Vulkan’s low-level nature can yield better performance for specific workloads, especially when both compute and graphics tasks are involved.
    \item \textbf{Ease of Use}: Vulkan is much harder to use than CUDA due to its low-level API and the need for explicit resource management.
\end{formatteditem}

\subsubsection{Vulkan Compute Code Example: Simple Compute Shader}

Here is a simple compute shader example in Vulkan for matrix addition:

\begin{verbatim}
// GLSL compute shader for Vulkan
#version 450
layout (local_size_x = 16, local_size_y = 16) in;

layout (binding = 0) buffer A { float a[]; };
layout (binding = 1) buffer B { float b[]; };
layout (binding = 2) buffer C { float c[]; };

void main() {
    uint idx = gl_GlobalInvocationID.x + gl_GlobalInvocationID.y * gl_WorkGroupSize.x;
    c[idx] = a[idx] + b[idx];
}
\end{verbatim}

This shader would be dispatched via a Vulkan compute pipeline, where resource and memory management must be manually handled.

\subsection{Metal: Apple's Proprietary GPU Programming Model}

Metal is Apple’s proprietary framework for GPU programming, aimed at macOS, iOS, and Apple Silicon hardware. It provides a unified framework for both graphics and GPGPU tasks, similar to Vulkan.

\subsubsection{Overview of Metal}

Metal is highly optimized for Apple’s ecosystem, providing developers with low-level access to the GPU, making it an ideal choice for applications on macOS and iOS. Metal supports both graphics and compute workloads, and it integrates deeply with Apple's hardware, including the M1 and M2 chips, which feature integrated GPUs.

\subsubsection{Metal vs. CUDA}

\begin{formatteditem}
    \item \textbf{Portability}: Metal is restricted to Apple hardware, whereas CUDA is restricted to NVIDIA hardware.
    \item \textbf{Performance}: Metal is highly optimized for Apple devices, and performance is excellent on that platform.
    \item \textbf{Ease of Use}: Metal offers a more streamlined API than Vulkan but is not as straightforward as CUDA for compute tasks.
\end{formatteditem}

\subsubsection{Metal Code Example: Simple Matrix Multiplication}

Below is a Metal example for matrix multiplication:

\begin{verbatim}
// Metal compute kernel for matrix multiplication
kernel void matrix_multiply(device float* A [[buffer(0)]],
                            device float* B [[buffer(1)]],
                            device float* C [[buffer(2)]],
                            uint id [[thread_position_in_grid]]) {
    C[id] = A[id] * B[id];
}

int main() {
    // Metal device and buffer setup
    id<MTLDevice> device = MTLCreateSystemDefaultDevice();
    id<MTLBuffer> bufferA = [device newBufferWithLength:size options:0];
    id<MTLBuffer> bufferB = [device newBufferWithLength:size options:0];
    id<MTLBuffer> bufferC = [device newBufferWithLength:size options:0];

    // Command queue and kernel dispatch
    id<MTLCommandQueue> commandQueue = [device newCommandQueue];
    id<MTLCommandBuffer> commandBuffer = [commandQueue commandBuffer];
    id<MTLComputeCommandEncoder> encoder = [commandBuffer computeCommandEncoder];
    [encoder setBuffer:bufferA offset:0 atIndex:0];
    [encoder setBuffer:bufferB offset:0 atIndex:1];
    [encoder setBuffer:bufferC offset:0 atIndex:2];
    [encoder dispatchThreads:MTLSizeMake(N, N, 1) threadsPerThreadgroup:MTLSizeMake(16, 16, 1)];
    [encoder endEncoding];
    [commandBuffer commit];
}
\end{verbatim}

\subsection{OpenGL: Compute Shaders for GPGPU}

OpenGL \cite{shreiner2009opengl}, traditionally known as a graphics API, can also perform GPGPU tasks through compute shaders. Compute shaders in OpenGL provide a means to perform parallel computation that is not directly related to graphics, using the same GPU pipeline. This makes OpenGL viable for GPGPU tasks when coupled with existing rendering operations, particularly in real-time applications.

\subsubsection{Overview of OpenGL Compute Shaders}

OpenGL compute shaders allow developers to write GPU programs for parallel computation. These shaders use GLSL (OpenGL Shading Language) and are designed to execute independent of the traditional graphics pipeline, making them useful for non-graphical computations such as physics simulations, particle systems, and data processing.

\subsubsection{OpenGL vs. CUDA}

OpenGL’s compute capabilities are not as extensive as CUDA or OpenCL, but for certain applications, they provide enough flexibility. OpenGL is primarily geared toward graphics tasks, while CUDA is specifically optimized for parallel computing on NVIDIA GPUs.

\begin{formatteditem}
    \item \textbf{Portability}: OpenGL is widely supported on different platforms and hardware, including NVIDIA, AMD, and Intel.
    \item \textbf{Performance}: Compute shaders in OpenGL are less optimized for raw compute tasks compared to CUDA or OpenCL, but performance can still be high in combined graphics-compute workflows.
    \item \textbf{Ease of Use}: OpenGL compute shaders can be more complex to use for GPGPU tasks due to its graphics-centric nature, though they offer flexibility in mixed workloads.
\end{formatteditem}

\subsubsection{OpenGL Code Example: Simple Compute Shader for Vector Addition}

Here’s a basic OpenGL compute shader that performs vector addition:

\begin{verbatim}
// GLSL compute shader for OpenGL
#version 430
layout (local_size_x = 16) in;

layout (std430, binding = 0) buffer A { float a[]; };
layout (std430, binding = 1) buffer B { float b[]; };
layout (std430, binding = 2) buffer C { float c[]; };

void main() {
    uint idx = gl_GlobalInvocationID.x;
    c[idx] = a[idx] + b[idx];
}
\end{verbatim}

The compute shader is dispatched in OpenGL like this:

\begin{verbatim}
// OpenGL code to dispatch the compute shader
GLuint computeShader = glCreateShader(GL_COMPUTE_SHADER);
glShaderSource(computeShader, 1, &shaderSource, NULL);
glCompileShader(computeShader);

GLuint program = glCreateProgram();
glAttachShader(program, computeShader);
glLinkProgram(program);
glUseProgram(program);

glDispatchCompute(1024 / 16, 1, 1);
glMemoryBarrier(GL_SHADER_STORAGE_BARRIER_BIT);
\end{verbatim}

This example demonstrates the use of compute shaders in OpenGL to perform vector addition. Although the API is primarily for graphics, compute shaders can be a powerful tool for non-graphical GPGPU tasks, especially when integrated with rendering pipelines.

\subsubsection{OpenGL vs. Vulkan Compute}

Both OpenGL and Vulkan support compute shaders, but Vulkan offers more fine-grained control over the GPU. OpenGL is simpler to set up but less flexible and less performant for compute tasks.

\begin{formatteditem}
    \item \textbf{Portability}: Both APIs are cross-platform, but Vulkan provides better control over hardware resources.
    \item \textbf{Performance}: Vulkan is more optimized for compute tasks, while OpenGL is typically used when combining rendering and compute.
    \item \textbf{Ease of Use}: OpenGL is easier to work with, but Vulkan offers more advanced features for compute.
\end{formatteditem}

\subsection{Conclusion On Other GPU Programming Models}

While CUDA remains the go-to for NVIDIA GPUs, alternatives like OpenCL, OpenGL, Vulkan Compute, and Metal provide viable options for cross-platform development. Each model has its strengths and weaknesses depending on the target hardware, complexity of the task, and specific performance requirements. Developers should choose the right programming model based on their use case, platform needs, and performance optimization goals.

\section{Conclusion}

Optimizing GPU algorithms requires careful attention to memory access patterns, shared memory utilization, and thread synchronization. By applying techniques such as memory coalescing, minimizing warp divergence, and using shared memory efficiently, we can achieve significant performance improvements in CUDA programs.

\chapter{Advanced CUDA Features and Optimization Techniques}

\section{Streams and Concurrency}
CUDA streams provide a mechanism to overlap computation and data transfer, allowing us to optimize the GPU's utilization by performing multiple tasks in parallel. The basic idea is that instead of serializing operations on the GPU (like launching one kernel and waiting for its completion), we can split tasks across multiple streams and execute them concurrently.

\subsection{Overlapping Computation and Data Transfer}
By default, CUDA operates in a synchronous manner: memory transfers (from host to device or device to host) and kernel executions are serialized, meaning one must finish before the other begins. However, we can overlap memory transfers with kernel execution using streams, which allows for more efficient use of the GPU.

To demonstrate overlapping, let's look at the following example.

\begin{lstlisting}[style=python]
import numpy as np
import pycuda.driver as cuda
import pycuda.autoinit
from pycuda.compiler import SourceModule

# Define the kernel
mod = SourceModule("""
__global__ void kernel(float *a, float *b) {
    int idx = threadIdx.x + blockIdx.x * blockDim.x;
    b[idx] = a[idx] * 2;
}
""")
kernel = mod.get_function("kernel")

# Initialize host data
N = 1024
h_a = np.random.randn(N).astype(np.float32)
h_b = np.empty_like(h_a)

# Allocate device memory
d_a = cuda.mem_alloc(h_a.nbytes)
d_b = cuda.mem_alloc(h_b.nbytes)

# Create streams
stream1 = cuda.Stream()
stream2 = cuda.Stream()

# Transfer data asynchronously in stream1
cuda.memcpy_htod_async(d_a, h_a, stream1)

# Launch kernel in stream2
kernel(d_a, d_b, block=(256, 1, 1), grid=(N // 256, 1), stream=stream2)

# Transfer result back asynchronously in stream1
cuda.memcpy_dtoh_async(h_b, d_b, stream1)

# Synchronize streams
stream1.synchronize()
stream2.synchronize()

# Output result
print(h_b[:10])
\end{lstlisting}

In this example:
- We created two streams: \texttt{stream1} and \texttt{stream2}.
- Data transfer (host to device and device to host) occurs in \texttt{stream1}, and kernel execution happens in \texttt{stream2}.
- By running these operations concurrently in different streams, we achieve better utilization of both the memory bandwidth and computational power of the GPU.

\subsection{Managing Multiple Streams}
Managing multiple streams becomes essential when optimizing more complex applications. CUDA streams are independent, and operations issued to different streams can be executed concurrently. However, there may be cases where we want to ensure the correct order of execution between streams.

To manage dependencies between streams, we can use \texttt{cudaStreamWaitEvent} to synchronize streams based on certain events. This ensures that kernels in one stream only start after a specific event in another stream has occurred.

Here’s a simple example:

\begin{lstlisting}[style=python]
# Creating an event
event = cuda.Event()

# Record event after memory copy in stream1
cuda.memcpy_htod_async(d_a, h_a, stream1)
event.record(stream1)

# Make stream2 wait for stream1 to finish
stream2.wait_for_event(event)

# Now we can safely execute the kernel in stream2
kernel(d_a, d_b, block=(256, 1, 1), grid=(N // 256, 1), stream=stream2)
\end{lstlisting}

In this code, we used an event to ensure that stream2 only starts its kernel after the data transfer in stream1 is complete.

\section{Dynamic Parallelism}
Dynamic parallelism in CUDA allows a kernel to launch other kernels directly from the device. This is useful for algorithms where the problem size is not known in advance or is irregular, such as adaptive mesh refinement, graph traversal, or recursive algorithms.

\subsection{Launching Kernels from within Kernels}
With dynamic parallelism, we can avoid returning to the CPU to launch new kernels. This significantly reduces the overhead and allows the GPU to make decisions dynamically.

Here’s a simple example of launching a kernel from within another kernel:

\begin{lstlisting}[style=python]
mod = SourceModule("""
__global__ void childKernel(int *data) {
    int idx = threadIdx.x + blockIdx.x * blockDim.x;
    data[idx] *= 2;
}

__global__ void parentKernel(int *data) {
    int idx = threadIdx.x + blockIdx.x * blockDim.x;

    // Launch child kernel if the condition is met
    if (idx == 0) {
        childKernel<<<1, 32>>>(data);
    }
}
""")
parentKernel = mod.get_function("parentKernel")

# Initialize host and device memory
data = np.array([1, 2, 3, 4, 5], dtype=np.int32)
d_data = cuda.mem_alloc(data.nbytes)
cuda.memcpy_htod(d_data, data)

# Launch parent kernel
parentKernel(d_data, block=(32, 1, 1), grid=(1, 1))

# Copy back result
cuda.memcpy_dtoh(data, d_data)
print(data)
\end{lstlisting}

Here, \texttt{parentKernel} launches \texttt{childKernel} directly from the GPU, showcasing dynamic parallelism.

\section{Profiling and Performance Tuning}
Performance tuning in CUDA involves identifying bottlenecks in both memory usage and computation, and then optimizing kernel execution. To achieve this, we can use profiling tools like \texttt{nvprof}, \texttt{Nsight Systems}, or \texttt{cuda-memcheck} to analyze our program.

\subsection{Using Profilers to Identify Bottlenecks}
Profilers provide insights into execution time, memory usage, and other performance metrics. By analyzing the output of these tools, we can pinpoint areas where our application is underperforming.

A common scenario is identifying memory transfer bottlenecks. For example, using \texttt{nvprof}:

\begin{lstlisting}[style=cmd]
$ nvprof python cuda_program.py
\end{lstlisting}

The profiler will show detailed information about the kernel's execution time, memory transfer time, and any potential bottlenecks in the code.

\subsection{Fine-Tuning Memory and Execution Strategies}
Memory is one of the most critical aspects of CUDA performance. Optimizing global, shared, and constant memory usage can significantly impact performance.

For example, if we notice that global memory access is slow, we can move frequently used data to shared memory, which is much faster. Here's an example of using shared memory to optimize a kernel:

\begin{lstlisting}[style=python]
mod = SourceModule("""
__global__ void kernelWithSharedMemory(float *a, float *b) {
    __shared__ float shared_data[256];

    int idx = threadIdx.x + blockIdx.x * blockDim.x;

    // Load data into shared memory
    shared_data[threadIdx.x] = a[idx];
    __syncthreads();

    // Perform computation
    b[idx] = shared_data[threadIdx.x] * 2;
}
""")
kernel = mod.get_function("kernelWithSharedMemory")

# Launch kernel
kernel(d_a, d_b, block=(256, 1, 1), grid=(N // 256, 1))
\end{lstlisting}

In this example, we use shared memory to store data that is frequently accessed by threads, reducing global memory access and improving performance.

By combining profiling tools and memory optimization strategies, we can fine-tune our CUDA applications for maximum performance.

\chapter{Applications of GPGPU in Modern Computing}

\section{High-Level GPU Libraries Overview}

As GPU computing has evolved, high-level libraries have emerged to simplify the development of complex applications such as deep learning \cite{peng2024deeplearningmachinelearning}, scientific computing, and data analytics. These libraries abstract many of the low-level details of CUDA and other GPU programming models, allowing developers to focus on algorithms and applications rather than GPU memory management and kernel execution.

In this section, we will provide a high-level overview of some of the most widely used GPU-accelerated libraries: \texttt{cuBLAS}, \texttt{cuDNN}, \texttt{TensorRT}, \texttt{PyTorch}, and \texttt{TensorFlow}. Each library is designed to accelerate specific types of operations, from linear algebra to deep neural networks, and they integrate seamlessly with CUDA to maximize performance on NVIDIA GPUs.

\subsection{cuBLAS: Accelerating Linear Algebra}

\texttt{cuBLAS} \cite{cui2011automatic} is an optimized GPU-accelerated library that implements standard BLAS (Basic Linear Algebra Subprograms) operations, such as matrix multiplication, vector addition, and solving linear systems. It is widely used in scientific computing and high-performance applications that require efficient numerical computations.

By using \texttt{cuBLAS}, developers can leverage the full power of NVIDIA GPUs without manually writing CUDA kernels. The library handles parallelization, memory transfers, and optimization, making it an essential tool for applications involving large-scale matrix operations.

\subsection{cuDNN: GPU-Optimized Neural Networks}

\texttt{cuDNN} (CUDA Deep Neural Network library) \cite{chetlur2014cudnn} is specifically designed to optimize the performance of deep learning algorithms on NVIDIA GPUs. It provides highly efficient implementations of key neural network operations, including convolutions, activation functions, pooling, and normalization layers.

\texttt{cuDNN} is integrated into popular deep learning frameworks like \texttt{PyTorch} and \texttt{TensorFlow}, significantly accelerating both the training and inference of neural networks. The library automatically optimizes operations to take full advantage of GPU hardware, making it ideal for deep learning tasks that require high throughput and low latency.

\subsection{TensorRT: Optimizing Inference for Deep Learning}

\texttt{TensorRT} \cite{jocher2022ultralytics} is an SDK developed by NVIDIA to optimize neural network inference on GPUs. It is particularly useful for deploying trained models in production environments where real-time performance is critical. \texttt{TensorRT} performs optimizations such as layer fusion, precision calibration (e.g., FP16 or INT8 quantization), and kernel auto-tuning to accelerate inference.

By using \texttt{TensorRT}, developers can take pre-trained models and optimize them for deployment, ensuring that neural networks run efficiently on a variety of NVIDIA GPUs in applications like autonomous vehicles, medical imaging, and recommendation systems.

\subsection{PyTorch: A Flexible Deep Learning Framework}

\texttt{PyTorch} is an open-source machine learning framework that has become highly popular in both research and industry. One of its key features is the dynamic computational graph, which allows for greater flexibility when designing and modifying neural networks. \texttt{PyTorch} also provides native support for CUDA, enabling seamless GPU acceleration for deep learning and tensor operations.

\texttt{PyTorch} is often preferred for research and experimentation due to its ease of use, flexible API, and strong community support. It is commonly used for training deep learning models on GPUs, with the added benefit of being easy to integrate with \texttt{cuDNN} for further performance optimizations.

\subsection{TensorFlow: An End-to-End Machine Learning Platform}

\texttt{TensorFlow}, developed by Google, is one of the most widely used machine learning platforms. It offers a comprehensive set of tools for building, training, and deploying machine learning models at scale. \texttt{TensorFlow} integrates with CUDA for GPU-accelerated training, making it highly efficient for large-scale machine-learning tasks.

While \texttt{TensorFlow} is known for its scalability and ability to run distributed workloads, it also provides high-level APIs (like \texttt{Keras}) for easier model development. \texttt{TensorFlow} is widely used in production environments, particularly for applications that require training and deploying neural networks on GPU clusters.

\subsection{Conclusion}

High-level GPU libraries such as \texttt{cuBLAS}, \texttt{cuDNN}, \texttt{TensorRT}, \texttt{PyTorch}, and \texttt{TensorFlow} offer powerful abstractions over low-level GPU programming models. These libraries allow developers to harness the computational power of GPUs with minimal effort, making them indispensable tools in fields like machine learning, scientific computing, and data analytics. Each library serves a specific purpose, and choosing the right one depends on the type of application and the level of control needed. 

As we move into the next section on GPGPU in Machine Learning, many of these libraries will be integral to understanding how deep learning workloads can be accelerated on GPUs.

\section{GPGPU in Machine Learning}

\subsection{Accelerating Neural Networks with GPUs}

GPUs (Graphics Processing Units) are designed to handle large-scale parallel computations, which makes them perfect for the task of training neural networks. While CPUs are optimized for general-purpose tasks, GPUs are specialized in performing the same operation across large amounts of data, which is common in machine learning tasks like matrix multiplications and tensor operations.

For example, consider a simple neural network training task where large batches of data are processed at once. A CPU might execute this task sequentially, while a GPU can parallelize it, leading to significant speedups.

Here’s an example of a basic neural network using \texttt{PyTorch}, a popular Python machine learning library, where we utilize GPU acceleration:

\begin{lstlisting}[style=python]
import torch
import torch.nn as nn
import torch.optim as optim

# Define a simple neural network
class SimpleNN(nn.Module):
    def __init__(self):
        super(SimpleNN, self).__init__()
        self.fc1 = nn.Linear(10, 50)
        self.fc2 = nn.Linear(50, 1)

    def forward(self, x):
        x = torch.relu(self.fc1(x))
        x = self.fc2(x)
        return x

# Create a model instance
model = SimpleNN()

# Check if GPU is available and move the model to GPU if it is
device = torch.device("cuda" if torch.cuda.is_available() else "cpu")
model.to(device)

# Generate random input and move it to the GPU
input_data = torch.randn(100, 10).to(device)

# Forward pass
output = model(input_data)
print(output)
\end{lstlisting}

In this example, we first check if a GPU is available using \texttt{torch.device("cuda")} and then move both the model and data to the GPU. This simple adjustment allows us to utilize GPU acceleration, which can make training much faster, especially with larger models and datasets.

\subsection{Tensor Operations on GPUs}

One of the key reasons why GPUs are so effective for machine learning is their ability to handle tensor operations efficiently. A tensor is a multi-dimensional array, and GPU-accelerated libraries like \texttt{PyTorch} and \texttt{TensorFlow} are designed to perform operations on these tensors in parallel.

Here’s a simple example that shows how to move a tensor operation to the GPU:

\begin{lstlisting}[style=python]
# Create a large tensor and move it to GPU
tensor = torch.randn(1000, 1000).to(device)

# Perform a matrix multiplication
result = torch.matmul(tensor, tensor)
print(result)
\end{lstlisting}

In this example, the \texttt{torch.matmul} function is performed on the GPU. Tensor operations like this are the foundation of many machine learning algorithms, including the backpropagation used to train neural networks.

\section{Scientific Computing and Simulations}

\subsection{Solving Large Linear Systems}

Scientific computing often involves solving large linear systems of equations, which can be computationally expensive on a CPU. However, these tasks can be parallelized and efficiently executed on a GPU.

For example, in Python, libraries such as \texttt{cuBLAS} and \texttt{cuSOLVER}, which are available through \texttt{PyCUDA} and \texttt{PyTorch}, allow you to solve large linear systems on the GPU.

Here’s an example of solving a linear system using \texttt{PyTorch} on a GPU:

\begin{lstlisting}[style=python]
# Create a large random matrix and a solution vector
A = torch.randn(1000, 1000).to(device)
b = torch.randn(1000).to(device)

# Solve the system of equations Ax = b
x = torch.linalg.solve(A, b)
print(x)
\end{lstlisting}

In this example, the \texttt{torch.linalg.solve} function solves the linear system $Ax = b$ on the GPU, where $A$ is a matrix and $b$ is a vector. This method is highly optimized and can significantly speed up scientific computations.

\subsection{GPU-Powered Simulations in Physics and Chemistry}

GPU acceleration is also widely used in simulations for physics and chemistry, such as molecular dynamics or finite element methods. These simulations involve solving differential equations or performing millions of small calculations simultaneously, which are well-suited for GPUs.

For instance, the popular simulation package \texttt{LAMMPS} supports GPU acceleration through CUDA, which allows researchers to simulate physical systems like materials and chemical reactions more efficiently.

\section{Real-Time Rendering and Graphics}

\subsection{GPU in Ray Tracing and Image Processing}

Real-time rendering tasks, such as ray tracing in computer graphics, rely heavily on GPU power to render scenes in real time. Ray tracing involves calculating the paths of rays of light through a scene, which is a highly parallelizable task and thus fits perfectly within a GPU’s architecture.

For example, NVIDIA’s RTX GPUs \cite{sanzharov2019examination} are specifically designed to accelerate ray tracing with hardware-accelerated ray tracing cores.

Here’s a simple example of how image processing can benefit from GPUs using \texttt{PyTorch}:

\begin{lstlisting}[style=python]
from PIL import Image
import torchvision.transforms as transforms

# Load an image
image = Image.open('sample.jpg')

# Convert image to tensor and move to GPU
image_tensor = transforms.ToTensor()(image).unsqueeze(0).to(device)

# Apply a simple filter (e.g., Gaussian blur)
blurred_image = torch.nn.functional.conv2d(image_tensor, weight)
\end{lstlisting}

In this example, we load an image, convert it into a tensor, and perform a convolution operation on the GPU, demonstrating the power of GPU acceleration in image processing tasks.

\section{Blockchain and Cryptocurrency Mining}

Bitcoin, the first decentralized cryptocurrency, was introduced in 2008 by an individual or group of individuals using the pseudonym \textit{Satoshi Nakamoto}. Nakamoto’s whitepaper, titled \textit{“Bitcoin: A Peer-to-Peer Electronic Cash System”} \cite{nakamoto2008bitcoin}, outlined a new form of digital currency that operated independently of traditional financial institutions, using cryptographic proof instead of trust. This marked the birth of blockchain technology, the underlying structure of Bitcoin.

In January 2009, Nakamoto mined the first block of the Bitcoin blockchain, known as the \textbf{genesis block} or \textbf{Block 0}. Embedded in this initial block was a message referencing a headline from \textit{The Times} newspaper on January 3, 2009: \textit{“Chancellor on brink of second bailout for banks”} \cite{duncan2009chancellor}. This message was widely interpreted as Nakamoto’s critique of the centralized banking system and the bailouts provided to financial institutions during the 2008 global financial crisis. Bitcoin, therefore, was born out of a vision to create a decentralized alternative to traditional currency, free from the control of central banks.

As Bitcoin began to gain traction, its decentralized and open nature allowed anyone with computational resources to participate in the network by mining new blocks. Mining not only helped secure the network but also rewarded participants with newly minted bitcoins. This led to the rapid growth of interest in Bitcoin, especially after its value began to rise significantly in the early 2010s. What started as a niche interest among cryptography enthusiasts quickly turned into a global phenomenon, with Bitcoin’s price reaching new heights and gaining mainstream attention.

Mining, initially performed by individual users with personal computers, soon evolved into a highly competitive and resource-intensive industry. As the value of Bitcoin surged, miners sought more efficient ways to solve the complex cryptographic puzzles that formed the basis of Bitcoin’s Proof of Work (PoW) consensus mechanism. This led to the development of specialized mining hardware, and mining became a race for computational power, with large-scale mining farms and mining pools dominating the landscape.

Today, Bitcoin and other cryptocurrencies have become an integral part of the global financial ecosystem, and the mining process continues to play a crucial role in maintaining the security and integrity of these decentralized networks.

\subsection{The Purpose of Mining and the Importance of Proof of Work (PoW)}

Cryptocurrency mining serves a crucial role in maintaining the integrity and security of decentralized blockchain networks. In the context of Bitcoin and other PoW-based cryptocurrencies, mining has two main purposes: validating and recording transactions on the blockchain and ensuring network security by preventing malicious attacks such as double-spending.

\textbf{Proof of Work (PoW)} \cite{kiayias2020proof} is the consensus algorithm used in many blockchain systems, like Bitcoin. In PoW, miners compete to solve complex cryptographic puzzles. The first miner to solve the puzzle gets the right to add a new block of transactions to the blockchain and is rewarded with cryptocurrency. This process is referred to as “mining.” 

The cryptographic puzzles involve hashing, and the goal is to find a hash output with a certain number of leading zeros, which makes the problem computationally expensive. The difficulty of the puzzle adjusts automatically to ensure that blocks are mined at regular intervals (e.g., every 10 minutes for Bitcoin).

\subsubsection{Why is PoW Important?}

PoW plays a critical role in securing the network through the following mechanisms:

\begin{itemize}
    \item \textbf{Prevents Double Spending:} Miners work to verify the legitimacy of transactions, ensuring that cryptocurrency is not spent more than once.
    \item \textbf{Decentralization:} PoW allows anyone with sufficient computational power to participate in the network, fostering decentralization by distributing control over the blockchain.
    \item \textbf{Economic Incentives:} By rewarding miners with newly created cryptocurrency and transaction fees, PoW incentivizes participants to act honestly and maintain the network's security.
    \item \textbf{Energy and Cost as Barriers:} The cost and energy required to perform PoW act as barriers to malicious actors. To alter a block in the blockchain, an attacker would need to redo the PoW for that block and every subsequent block, which becomes practically infeasible as the chain grows.
\end{itemize}

Without PoW, blockchain networks would be vulnerable to attacks, including Sybil attacks or attempts to rewrite transaction history.

\subsection{The Mining Arms Race}

As cryptocurrency mining became more competitive and lucrative, miners began searching for ways to increase their computational power and improve their chances of successfully solving PoW puzzles. This led to a rapid evolution in mining hardware, often referred to as a “mining arms race.”

In the early days of Bitcoin, miners used \textbf{CPUs} (central processing units) to mine blocks. However, as the network grew and the difficulty of PoW puzzles increased, CPUs were quickly replaced by more powerful hardware:

\begin{itemize}
    \item \textbf{GPUs (Graphics Processing Units):} Due to their parallel processing capabilities, GPUs proved to be much more efficient than CPUs for mining, allowing miners to compute many hashes simultaneously.
    \item \textbf{FPGA (Field-Programmable Gate Arrays):} These devices offered even better performance and efficiency, as they could be tailored specifically for mining tasks.
    \item \textbf{ASICs (Application-Specific Integrated Circuits):} The ultimate evolution of mining hardware, ASICs are specialized chips designed exclusively for cryptocurrency mining. They are far more powerful and energy-efficient than both GPUs and FPGAs, but their development and production require significant financial investment.
\end{itemize}

\textbf{Mining Equipment Arms Race:} The shift from CPUs to GPUs and eventually ASICs marked the beginning of a hardware arms race. As mining became more profitable, large mining farms and pools emerged, often in regions with cheap electricity. These industrial-scale operations invest heavily in the latest ASIC technology to maintain a competitive edge, leading to a continuous cycle of hardware upgrades.

This arms race has made it difficult for individual miners with lower computational resources to compete, pushing many towards joining mining pools where they can combine their hashing power with others to increase their chances of receiving mining rewards.

\subsubsection{Impact of the Mining Arms Race}

The mining arms race has led to several key outcomes:
\begin{itemize}
    \item \textbf{Centralization of Mining:} While blockchain networks are designed to be decentralized, the high cost and efficiency of ASICs have concentrated mining power in the hands of a few large players, particularly mining pools.
    \item \textbf{Increased Energy Consumption:} As miners invest in more powerful hardware, energy consumption has risen sharply, leading to concerns about the environmental impact of large-scale mining operations.
    \item \textbf{Constant Upgrades:} The competitiveness of mining means that even ASIC miners must regularly upgrade their hardware to remain profitable, further intensifying the arms race.
\end{itemize}

\subsection{How GPUs Dominate Blockchain Computing}

Blockchain technologies, especially in the context of cryptocurrency mining, have seen a major boom due to the computational power of GPUs. In mining cryptocurrencies like Bitcoin or Ethereum, the GPU performs a repetitive task known as hashing to solve cryptographic puzzles.

GPUs are much faster than CPUs for this task because the problem can be divided into smaller parts and computed in parallel. While CPUs can perform a few tasks at once, GPUs are optimized for executing thousands of small, parallel operations simultaneously, which makes them ideal for mining.

The primary computational task in cryptocurrency mining is solving the cryptographic hash functions. A hash is a one-way function that maps input data of any size to a fixed-size output. The process involves finding a hash output with certain properties (such as a specific number of leading zeros in the case of Bitcoin), which requires generating many random inputs and checking the output hash values.

Here’s a simplified example of using Python to perform a cryptographic hash calculation on a CPU:

\begin{lstlisting}[style=python]
import hashlib

# Define a simple hash function
def calculate_hash(input_string):
    return hashlib.sha256(input_string.encode()).hexdigest()

# Test the hash function
hash_value = calculate_hash("Hello, World!")
print(hash_value)
\end{lstlisting}

In real blockchain mining, this process involves millions of hash calculations every second, which is why GPUs with their parallel processing capabilities are preferred. Libraries such as \texttt{Hashcat} or \texttt{NiceHash} are often used to mine cryptocurrencies, leveraging the GPU’s ability to compute hashes at high speed.

\subsection{GPU-Based Hashing Example}

Let’s consider how GPUs perform parallel computations using Python with a library like \texttt{NumPy}. Although this example is simplified compared to real-world mining, it demonstrates the parallel processing capabilities of GPUs.

\begin{lstlisting}[style=python]
import numpy as np
import hashlib
from numba import vectorize, cuda

# Define a hash function that will be executed in parallel on the GPU
@vectorize(['uint64(uint64)'], target='cuda')
def gpu_hash(input_value):
    input_str = str(input_value).encode()
    hash_output = hashlib.sha256(input_str).hexdigest()
    return int(hash_output, 16) & 0xffffffffffffffff  # Return 64-bit hash

# Create a large array of random numbers as inputs
input_values = np.random.randint(1, 1000000, size=1000000, dtype=np.uint64)

# Compute hashes in parallel using GPU
output_hashes = gpu_hash(input_values)

print(output_hashes[:10])  # Display first 10 hashed values
\end{lstlisting}

In this example, the GPU computes hashes in parallel for multiple input values. This is a simplified demonstration, but in real mining, GPUs perform similar tasks at much higher speeds and scales. Modern mining operations involve mining pools and dedicated hardware, such as ASICs (Application-Specific Integrated Circuits), but GPUs still play a critical role, especially in mining cryptocurrencies like Ethereum.

\section{GPU Virtualization and Cloud Computing}

As GPGPU workloads grow in size and complexity, cloud computing has become an essential tool for scaling computational resources on demand. Cloud platforms such as AWS, Google Cloud, and Microsoft Azure provide access to powerful GPU instances, allowing users to perform intensive computations without the need to invest in physical hardware. At the core of many cloud-based GPU services is \textbf{GPU virtualization}, a technique that allows multiple users or workloads to share a single GPU while maintaining isolation and performance.

This section explores the concept of GPU virtualization, its role in cloud computing, and how it enables scalable workloads. We will also discuss some key cloud platforms and provide insights into how to efficiently leverage GPU resources in the cloud.

\subsection{What is GPU Virtualization?}

GPU virtualization allows multiple virtual machines (VMs) or containers to share a single physical GPU, much like how CPU virtualization enables multiple VMs to share a single CPU. With GPU virtualization, cloud providers can offer GPU resources more efficiently, enabling multiple users to run workloads simultaneously on the same GPU.

There are different types of GPU virtualization techniques, including:

\begin{formatteditem}
    \item \textbf{Full GPU Passthrough}: In this model, the entire GPU is allocated to a single VM or container. This provides the best performance but lacks resource sharing.
    \item \textbf{Shared GPU (vGPU)}: In this model, a single physical GPU is split into multiple virtual GPUs (vGPUs), each assigned to different VMs or containers. This allows for efficient resource sharing and improved utilization.
    \item \textbf{Multi-Instance GPU (MIG)}: Introduced by NVIDIA, this technique allows physical GPUs to be partitioned into smaller, isolated instances. Each instance can handle separate workloads independently, improving resource efficiency.
\end{formatteditem}

GPU virtualization enables cloud providers to serve multiple customers on the same physical GPU, which increases scalability, reduces costs, and makes GPU resources more accessible to smaller organizations.

\subsection{GPU in Cloud Platforms}

Several cloud providers offer GPU instances optimized for machine learning, high-performance computing (HPC), and other GPU-accelerated workloads. Here’s an overview of how leading cloud platforms integrate GPUs into their offerings:

\begin{formatteditem}
    \item \textbf{Amazon Web Services (AWS)}: AWS offers EC2 instances with NVIDIA Tesla and A100 GPUs through the \texttt{p3} and \texttt{p4} instances. These instances are designed for large-scale machine learning and deep learning workloads.
    \item \textbf{Google Cloud Platform (GCP)}: Google Cloud provides GPU instances with NVIDIA Tesla and A100 hardware, which can be used for machine learning tasks, video rendering, and simulations.
    \item \textbf{Microsoft Azure}: Azure provides a variety of GPU-enabled virtual machines, including instances with NVIDIA V100 and A100 GPUs. These are used for AI, machine learning, and data analytics tasks.
\end{formatteditem}

Cloud platforms make it easy to scale GPU resources based on demand. Users can launch multiple GPU instances, distribute workloads across them, and terminate instances when tasks are complete, ensuring cost-effectiveness and flexibility.

\subsection{Benefits of GPU Virtualization and Cloud Computing}

GPU virtualization and cloud computing offer several advantages over traditional on-premises GPU deployment:

\begin{formatteditem}
    \item \textbf{Scalability}: Cloud platforms enable users to scale GPU resources up or down based on workload requirements. This is particularly useful for machine learning tasks where the demand for resources can vary over time.
    \item \textbf{Cost Efficiency}: By leveraging shared GPU resources through virtualization, users only pay for the GPU capacity they use. This eliminates the need to invest in expensive hardware upfront.
    \item \textbf{Flexibility}: Cloud-based GPUs provide access to a wide range of hardware configurations, allowing users to choose the most suitable setup for their specific applications.
    \item \textbf{Multi-Tenancy}: GPU virtualization allows multiple users or workloads to run on the same hardware simultaneously, improving overall GPU utilization and reducing idle resources.
\end{formatteditem}

\subsection{Running GPU Workloads in the Cloud: Example on AWS}

Let’s look at how you can run GPU-accelerated machine learning workloads on AWS using a GPU instance. Below is a simple example of setting up an AWS EC2 instance with a GPU to train a neural network:

\subsubsection{Steps to Launch an EC2 GPU Instance}

\begin{formatteditem}
    \item Go to the AWS EC2 dashboard and click \texttt{Launch Instance}.
    \item Select an AMI (Amazon Machine Image) that includes deep learning frameworks such as the \texttt{Deep Learning AMI}.
    \item Choose an instance type with GPU support, such as \texttt{p3.2xlarge}, which includes an NVIDIA Tesla V100 GPU.
    \item Configure the instance and launch it.
    \item Connect to the instance via SSH and ensure that CUDA and cuDNN are installed and configured correctly.
\end{formatteditem}
% TODO Add pictures for this

\subsection{GPU Virtualization for Multi-Tenant Workloads}

NVIDIA’s Multi-Instance GPU (MIG) technology allows cloud providers to partition GPUs into multiple instances, each capable of running separate workloads. This makes GPU resources more flexible and efficient, especially for multi-tenant environments.

For example, an A100 GPU with MIG can be split into up to seven independent GPU instances, each isolated from the others. This allows multiple users to run different tasks on the same physical GPU, maximizing hardware utilization while maintaining strong performance.

\subsection{Conclusion}

GPU virtualization and cloud computing have revolutionized the way organizations utilize GPU resources. By enabling scalable, flexible, and cost-efficient access to GPUs, cloud platforms make it easier than ever to run GPU-accelerated workloads, from machine learning to high-performance computing. GPU virtualization technologies, such as NVIDIA's Multi-Instance GPU, ensure that these resources can be shared among multiple users without compromising performance, making the cloud an attractive solution for businesses and researchers alike.

\chapter{Future of GPGPU and Emerging Trends}

\section{AI and GPUs: The Next Frontier}

GPUs (Graphics Processing Units) have become indispensable in the field of Artificial Intelligence (AI). This is largely because AI, especially machine learning and deep learning \cite{chen2024deeplearningmachinelearning}, requires large-scale parallel computations, something GPUs excel at. 

\textbf{Why GPUs for AI?}

GPUs are designed to handle thousands of operations simultaneously. This parallel processing capability makes them well-suited for AI tasks such as:

\begin{formatteditem}
    \item \textbf{Training Neural Networks:} During the training process, neural networks need to adjust millions of parameters (weights). GPUs speed this up by performing multiple calculations simultaneously.
    \item \textbf{Running Inference:} Once a model is trained, it needs to make predictions quickly. GPUs allow the model to run efficiently in real-time applications, such as self-driving cars or voice recognition systems.
\end{formatteditem}

\textbf{Example: Training a Simple Neural Network with a GPU}

In Python, using libraries like TensorFlow or PyTorch, you can easily utilize a GPU. Below is an example of how you might set up TensorFlow to use the GPU:

\begin{lstlisting}[style=python]
import tensorflow as tf

# Check if a GPU is available
print("Num GPUs Available: ", len(tf.config.list_physical_devices('GPU')))

# Define a simple neural network
model = tf.keras.models.Sequential([
    tf.keras.layers.Dense(128, activation='relu', input_shape=(784,)),
    tf.keras.layers.Dense(10, activation='softmax')
])

# Compile the model
model.compile(optimizer='adam', loss='sparse_categorical_crossentropy', metrics=['accuracy'])

# Train the model
model.fit(training_data, training_labels, epochs=5)
\end{lstlisting}

This code automatically uses the GPU if one is available, providing a significant performance boost when training the neural network.

\section{Integration of GPUs with Other Processing Units}

While GPUs are powerful for parallel computations, CPUs (Central Processing Units) still excel at general-purpose tasks and sequential processing. Combining these two processors can create a highly efficient system.

\subsection{CPU-GPU Synergy}

A CPU is like the "brain" of the computer, handling complex logic, but it cannot perform parallel operations as quickly as a GPU. Combining the strengths of both allows for a system where the CPU can handle high-level control while offloading computationally intensive tasks to the GPU.

\textbf{Example: CPU-GPU Workflow in Python}

In Python, libraries like \texttt{cuPy} allow you to write code that seamlessly transfers data between the CPU and GPU. Here's a simple example of multiplying matrices using both a CPU and a GPU.

\begin{lstlisting}[style=python]
import numpy as np
import cupy as cp

# CPU computation
a_cpu = np.random.rand(10000, 10000)
b_cpu = np.random.rand(10000, 10000)

# Multiply matrices on the CPU
result_cpu = np.dot(a_cpu, b_cpu)

# GPU computation
a_gpu = cp.random.rand(10000, 10000)
b_gpu = cp.random.rand(10000, 10000)

# Multiply matrices on the GPU
result_gpu = cp.dot(a_gpu, b_gpu)
\end{lstlisting}

In this example, \texttt{cuPy} allows us to leverage the power of the GPU for matrix multiplication, which is significantly faster than performing the same operation on the CPU.

\subsection{GPU and FPGA Hybrid Systems}

FPGAs (Field Programmable Gate Arrays) are hardware circuits that can be configured after manufacturing. Combining GPUs and FPGAs can offer significant advantages in specialized computing tasks such as cryptography, AI inference, and high-frequency trading.

\textbf{Example: GPU and FPGA Integration in a Computing System}

Consider a system where the FPGA handles specialized data preprocessing tasks, and the GPU handles the bulk of machine learning computations. Below is a simple representation of this hybrid architecture:

\begin{itemize}
    \item \textbf{Input}
        \begin{itemize}
            \item Data is first processed by an \textbf{FPGA} (Field Programmable Gate Array), which specializes in handling specific tasks like data preprocessing.
        \end{itemize}
    \item After the data preprocessing is complete, it is passed to the \textbf{GPU} (Graphics Processing Unit).
        \begin{itemize}
            \item The GPU performs heavy computational tasks, such as running machine learning models or deep learning training.
        \end{itemize}
    \item Finally, the processed data is passed to the \textbf{Output} stage.
\end{itemize}

This hybrid system allows each processing unit to focus on what it does best, improving overall performance and efficiency.

\section{Quantum Computing and GPUs: What’s Next?}

Quantum computing is still in its early stages but holds great promise for solving certain types of problems that are currently too complex for classical computers, including GPUs. While GPUs are excellent for parallel tasks, quantum computers may one day perform computations that GPUs cannot handle efficiently.

\textbf{How Will Quantum Computing Affect GPUs?}

Quantum computers and GPUs may complement each other. For example:
- \textbf{Simulating Quantum Algorithms:} Before real quantum computers become widely available, GPUs can be used to simulate quantum algorithms.
- \textbf{Quantum-GPU Hybrid Systems:} In the future, hybrid systems may be developed where GPUs handle classical parallel computations and quantum computers handle quantum-specific tasks.

\textbf{Example: Simulating a Quantum Circuit on a GPU}

Below is an example of how you might simulate a simple quantum circuit using a Python library like Qiskit:

\begin{lstlisting}[style=python]
from qiskit import QuantumCircuit, Aer, execute

# Create a simple quantum circuit with 2 qubits
qc = QuantumCircuit(2)
qc.h(0)  # Apply Hadamard gate on qubit 0
qc.cx(0, 1)  # Apply CNOT gate between qubit 0 and 1

# Simulate the circuit on a classical computer (using GPU if available)
backend = Aer.get_backend('statevector_simulator')
job = execute(qc, backend)
result = job.result()

# Output the result
print(result.get_statevector())
\end{lstlisting}

When we run a quantum circuit like the above, the output is typically represented as a \textbf{state vector}, which describes the quantum state of the system at that point. In this case, we are working with a simple two-qubit system where we apply a \textbf{Hadamard gate} \cite{renner2022computational} to the first qubit and a \textbf{CNOT gate} \cite{bremner2002practical} to entangle the two qubits. Let's break this down step-by-step.

\textbf{Step 1: Initial state}

At the very start, both qubits are in the state \(|0\rangle\), which in vector form is represented as:

\[
|0\rangle = 
\begin{bmatrix}
1 \\
0 
\end{bmatrix}
\]

For a two-qubit system, the state \(|00\rangle\) is represented as:

\[
|00\rangle = 
\begin{bmatrix}
1 \\
0 \\
0 \\
0
\end{bmatrix}
\]

\textbf{Step 2: Applying the Hadamard gate}

The Hadamard gate creates superposition. When applied to the first qubit, it transforms \(|0\rangle\) into \(\frac{1}{\sqrt{2}}(|0\rangle + |1\rangle)\). So, the new state of the system becomes:

\[
\frac{1}{\sqrt{2}} \left( |00\rangle + |10\rangle \right) = 
\frac{1}{\sqrt{2}} 
\begin{bmatrix}
1 \\
0 \\
1 \\
0
\end{bmatrix}
\]

This means the system is now in an equal superposition of the states \(|00\rangle\) and \(|10\rangle\).

\textbf{Step 3: Applying the CNOT gate}

The CNOT gate entangles the qubits. It flips the second qubit only when the first qubit is in the state \(|1\rangle\). Applying the CNOT gate \cite{bremner2002practical} to our system results in the state \(\frac{1}{\sqrt{2}}(|00\rangle + |11\rangle)\). In vector form, this looks like:

\[
\frac{1}{\sqrt{2}} 
\begin{bmatrix}
1 \\
0 \\
0 \\
1
\end{bmatrix}
\]

\textbf{Final result: Bell State}

This state is known as a \textbf{Bell state} or \textbf{EPR pair}, which is one of the most fundamental forms of quantum entanglement. The state \(\frac{1}{\sqrt{2}}(|00\rangle + |11\rangle)\) means that both qubits are perfectly correlated—if you measure the first qubit and find it in state \(|0\rangle\), the second qubit will also be in state \(|0\rangle\), and if the first qubit is in state \(|1\rangle\), the second qubit will also be in state \(|1\rangle\).

\textbf{State vector output:}

After running the quantum circuit simulation, you will get the following state vector as the output:

\[
\begin{bmatrix}
0.70710678+0.j \\
0.        +0.j \\
0.        +0.j \\
0.70710678+0.j
\end{bmatrix}
\]

This output represents the quantum state \(\frac{1}{\sqrt{2}}(|00\rangle + |11\rangle)\), confirming that the two qubits are now entangled.

While this code runs on a classical computer (and can utilize GPU acceleration for the simulation), future advancements might see such tasks delegated to real quantum processors.

In conclusion, the future of GPUs is exciting, especially in the context of AI, hybrid processing units, and the potential synergy with quantum computing. As technology evolves, understanding these trends will be crucial for developers.

\chapter{Take it Easy!}

\section{Introduction}

In the previous chapter, we delved deep into CUDA and how to optimize C++ code for better performance on GPUs. While understanding the low-level details can be valuable, it’s important to know that you don’t have to master these techniques to leverage the power of deep learning in practice. Many of the algorithms, optimizations, and matrix operations we explored are already implemented and highly optimized in libraries like PyTorch. 

When building deep learning models, whether you're working with simple feedforward neural networks or more complex convolutional neural networks (CNNs), PyTorch provides all the tools you need, abstracting away the need to write raw CUDA code. This allows you to focus more on the logic and design of your neural networks rather than the underlying hardware optimizations.

In this chapter, we will demonstrate how easy it is to create, train, and optimize neural networks using PyTorch. We'll guide you through:

\begin{enumerate}
    \item A simple dense neural network for the MNIST dataset.
    \item A convolutional neural network (CNN) for the CIFAR-10 dataset.
    \item Using the CLIP model to extract image features.
\end{enumerate}

By the end of this chapter, you’ll see how powerful and convenient it is to use high-level libraries like PyTorch for implementing deep learning models.

\section{Example 1: MNIST Dense Neural Network}

Let’s start with a simple fully connected neural network to classify handwritten digits from the MNIST dataset. In this example, we’ll build a dense network with one hidden layer.

\begin{lstlisting}[style=python]
import torch
import torch.nn as nn
import torch.optim as optim
import torchvision
import torchvision.transforms as transforms

# Define the neural network
class SimpleNN(nn.Module):
    def __init__(self):
        super(SimpleNN, self).__init__()
        self.fc1 = nn.Linear(28 * 28, 128)  # First fully connected layer
        self.fc2 = nn.Linear(128, 10)       # Output layer (10 classes)

    def forward(self, x):
        x = x.view(-1, 28 * 28)  # Flatten the input
        x = torch.relu(self.fc1(x))
        x = self.fc2(x)
        return x

# Load the MNIST dataset
transform = transforms.Compose([transforms.ToTensor(), transforms.Normalize((0.5,), (0.5,))])
trainset = torchvision.datasets.MNIST(root='./data', train=True, download=True, transform=transform)
trainloader = torch.utils.data.DataLoader(trainset, batch_size=64, shuffle=True)

# Instantiate the model, define the loss function and the optimizer
model = SimpleNN()
criterion = nn.CrossEntropyLoss()
optimizer = optim.Adam(model.parameters(), lr=0.001)

# Training loop
for epoch in range(5):  # Train for 5 epochs
    running_loss = 0.0
    for images, labels in trainloader:
        optimizer.zero_grad()  # Zero the gradients
        outputs = model(images)  # Forward pass
        loss = criterion(outputs, labels)  # Compute loss
        loss.backward()  # Backpropagation
        optimizer.step()  # Update weights
        running_loss += loss.item()

    print(f'Epoch {epoch+1}, Loss: {running_loss/len(trainloader)}')

print("Finished Training")
\end{lstlisting}

In this example, you can see how easy it is to set up a simple neural network in PyTorch. The library takes care of many things for you, such as handling data, initializing layers, and optimizing parameters.

\section{Example 2: CIFAR-10 Convolutional Neural Network}

Next, let’s build a convolutional neural network (CNN) to classify images from the CIFAR-10 dataset. CNNs are more effective than dense networks for image data because they can capture spatial hierarchies in images.

\begin{lstlisting}[style=python]
import torch
import torch.nn as nn
import torch.optim as optim
import torchvision
import torchvision.transforms as transforms

# Define the CNN
class CNN(nn.Module):
    def __init__(self):
        super(CNN, self).__init__()
        self.conv1 = nn.Conv2d(3, 32, 3, padding=1)  # First conv layer
        self.pool = nn.MaxPool2d(2, 2)  # Max pooling
        self.conv2 = nn.Conv2d(32, 64, 3, padding=1)  # Second conv layer
        self.fc1 = nn.Linear(64 * 8 * 8, 512)  # Fully connected layer
        self.fc2 = nn.Linear(512, 10)          # Output layer (10 classes)

    def forward(self, x):
        x = self.pool(torch.relu(self.conv1(x)))
        x = self.pool(torch.relu(self.conv2(x)))
        x = x.view(-1, 64 * 8 * 8)  # Flatten for fully connected layer
        x = torch.relu(self.fc1(x))
        x = self.fc2(x)
        return x

# Load the CIFAR-10 dataset
transform = transforms.Compose([transforms.ToTensor(), transforms.Normalize((0.5, 0.5, 0.5), (0.5, 0.5, 0.5))])
trainset = torchvision.datasets.CIFAR10(root='./data', train=True, download=True, transform=transform)
trainloader = torch.utils.data.DataLoader(trainset, batch_size=64, shuffle=True)

# Instantiate the model, define the loss function and the optimizer
model = CNN()
criterion = nn.CrossEntropyLoss()
optimizer = optim.Adam(model.parameters(), lr=0.001)

# Training loop
for epoch in range(10):  # Train for 10 epochs
    running_loss = 0.0
    for images, labels in trainloader:
        optimizer.zero_grad()  # Zero the gradients
        outputs = model(images)  # Forward pass
        loss = criterion(outputs, labels)  # Compute loss
        loss.backward()  # Backpropagation
        optimizer.step()  # Update weights
        running_loss += loss.item()

    print(f'Epoch {epoch+1}, Loss: {running_loss/len(trainloader)}')

print("Finished Training")
\end{lstlisting}

In this CNN example, PyTorch handles the convolutional layers and max-pooling operations. Again, you don’t need to worry about how CUDA optimizes these computations under the hood. 

\section{Example 3: Using CLIP for Image Feature Extraction}

CLIP (Contrastive Language-Image Pretraining) is a model developed by OpenAI that can be used to extract features from images. It has been trained on a vast amount of data and can generate high-quality image representations.

Here’s a simple example of how to use CLIP to extract features from an image:

\begin{lstlisting}[style=python]
import torch
import clip
from PIL import Image

# Load the CLIP model
device = "cuda" if torch.cuda.is_available() else "cpu"
model, preprocess = clip.load("ViT-B/32", device=device)

# Preprocess the image
image = Image.open("example.jpg")
image_input = preprocess(image).unsqueeze(0).to(device)

# Extract image features
with torch.no_grad():
    image_features = model.encode_image(image_input)

print("Image features extracted:", image_features.shape)
\end{lstlisting}

In this example, we use the pre-trained CLIP model to generate feature vectors for an image. This can be useful in tasks like image retrieval or zero-shot classification.

\section{Conclusion}

As we’ve demonstrated in this chapter, PyTorch abstracts away many of the complexities associated with deep learning and GPU programming. Whether you're building a simple dense neural network for MNIST, a convolutional network for CIFAR-10, or leveraging advanced models like CLIP, you can do so without needing to dive into CUDA or C++.

The main takeaway is this: You don’t need to worry about the low-level details to get started with deep learning. Libraries like PyTorch are designed to handle all of that for you. So, relax, and let PyTorch do the heavy lifting!

\bibliographystyle{ieeetr}
% \bibliography{sample}
\bibliography{reference}

\end{document}